\documentclass[pra,a4paper,twocolumn,superscriptaddress,longbibliography,nofootinbib]{revtex4-2}

\usepackage{amssymb}
\usepackage{amsmath}
\usepackage{amsfonts}
\usepackage{graphicx}
\usepackage{bm}
\usepackage{xcolor}
\usepackage{multirow}
\usepackage{natbib}

\DeclareMathOperator{\Tr}{Tr}
\DeclareMathOperator{\spp}{sp}
\DeclareMathOperator{\sgn}{sgn}

\DeclareMathOperator{\re}{Re}
\DeclareMathOperator{\arcsinh}{arcsinh}

\begin{document}

\title{Multifractally-enhanced superconductivity in two-dimensional systems with spin-orbit coupling}

\author{E.~S. Andriyakhina}

\affiliation{Moscow Institute for Physics and Technology, 141700 Moscow, Russia}

\affiliation{\hbox{L.~D.~Landau Institute for Theoretical Physics, acad. Semenova av. 1-a, 142432 Chernogolovka, Russia}}

    \author{I.~S. Burmistrov}
\affiliation{\hbox{L.~D.~Landau Institute for Theoretical Physics, acad. Semenova av. 1-a, 142432 Chernogolovka, Russia}}

\affiliation{Laboratory for Condensed Matter Physics, HSE University, 101000 Moscow, Russia
}


\begin{abstract}The interplay
of Anderson localization and electron-electron interactions is known to lead to enhancement of superconductivity due to 
multifractality of electron wave functions. We develop the theory of multifractally-enhanced superconducting states in two-dimensional systems in the presence of spin-orbit coupling. Using the Finkel'stein nonlinear sigma model, we derive the modified Usadel and gap equations that take into account renormalizations caused by
the interplay of disorder and interactions. Multifractal correlations induce energy dependence of the superconducting spectral gap.
We determine the superconducting transition temperature and the superconducting spectral gap in the case of Ising and strong spin orbit couplings. In the latter case the energy dependence of superconducting spectral gap is convex whereas in the former case (as well as in the absence of spin-orbit coupling) it is concave.  Multifractality enhances not only the transition temperature but, in the same way, the spectral gap at zero temperature. Also we study mesoscopic fluctuations of the local density of states in the superconducting state. Similarly to the case of normal metal, spin-orbit coupling reduce the amplitude of fluctuations. 
\end{abstract}

\maketitle

\section{Introduction}

Superconductivity and Anderson localization are two fundamental quantum phenomena that have been still attracting great interest. Initially it was believed that non-magnetic disorder does not affect s-wave superconducting order parameter (so-called ``Anderson theorem'' \cite{Intro2,Intro3,Intro4}). Later the paradigm was shifted to consider superconductivity and disorder as antagonists due to Anderson localization \cite{AndersonLoc}. Strong localization 
were predicted to suppress superconductivity \cite{Intro5,Intro6,Intro7,Intro8}. Similar destruction of superconductivity was predicted due to Coulomb
interaction at weak disorder \cite{Intro9,Intro10,Intro11,Intro12,Intro13,Intro14,Intro15}. 
The experimental discovery \cite{Intro16i}  of superconductor-to-insulator transition boosted interest to effects of disorder on superconducting correlations in thin films (see Refs. \cite{Intro17i,Intro18i,Intro19i} for a review).

Recently, the paradigm has been shifted once again. It was predicted in Refs. \cite{Intro20i,Intro21i} that Anderson localization can lead to enhancement of the superconducting transition temperature, $T_c$ for systems near the Anderson transition (e.g., in three dimensions). The mechanism is based on the multifractal behavior of wave functions 
--- the well-known companion of Anderson localization --- that leads to  enhancement of effective attraction between electrons. This mechanism works in the absence of the long-ranged
Coulomb repulsion. Later, the multifractal enhancement of
$T_c$ has been predicted for systems in the regime of weak localization (or anti-localization) which is relevant for weakly disordered superconducting films 
\cite{BurmistrovPRL2012,BurmistrovPRB2015}. These analytical predictions have been further tested by numerical computations for the disordered attractive Hubbard model on a two-dimensional lattice \cite{Intro24,Intro33,Intro34}.
Recently, a growth of $T_c$  with increase of disorder observed in monolayer niobium dichalcogenides \cite{Intro25,Intro26} has been suggested as a demonstration of the multifractal-enhancement mechanism.

One way to characterize multifractally-enhanced superconducting state is to study the mesoscopic fluctuations of the local density of states \cite{BurmistrovAnnPhys2021,Stosiek2021}. Potentially, it can be very promising due to (i) many reported tunneling spectroscopy data of the point-to-point fluctuations of the local density of states in thin superconducting films \cite{Intro27,Intro28,Intro29,Intro30,Intro31,Intro32} and (ii) a qualitative agreement between the theory \cite{BurmistrovPRB2016} developed for temperatures $T>T_c$  and experiments on the local density of states in the normal phase of disordered superconducting films.

However, there exist superconducting thin films and two-dimensional systems with broken spin rotational symmetry due to the presence of spin-orbit coupling. Among them one can 
list single atomic layers of Pb on Si \cite{Brun2014}, SrTiO$_3$ surfaces \cite{Intro18,Intro19}, LaAlO$_3$/SrTiO$_3$ interfaces
\cite{Intro16,Intro17}, and MoS$_2$ flakes \cite{Intro20a,Intro20b,Intro21}. Moreover, 
spin orbit coupling of Ising type is expected to exist in monolayer niobium dichalcogenides in which multifractal enhancement of superconductivity was measured \cite{Intro25,Intro26}. This calls for the theory of multifractally-enhanced superconductivity in two-dimensional systems with spin orbit coupling.

In this work we extend the theory of the multifractal superconducting state  
developed in Ref. \cite{BurmistrovAnnPhys2021} to thin films with spin-orbit coupling. 
As in Ref. \cite{BurmistrovAnnPhys2021} we focus on the case of weak short-ranged electron-electron interaction \footnote{A long-range component (Coulomb) of interaction can be suppressed in films covered by a substrate with a high dielectric
constant.}. We consider an intermediate disorder strength (but still weak) in which analysis of the renormalization group equations for the normal state predicted parametrically enhanced $T_c$ in comparison with 
the conventional Bardeen-Cooper-Schrieffer (BCS) result \cite{BurmistrovPRL2012}. Using the Finkel'stein nonlinear sigma model, we derive the Usadel equation together with the equation for the spectral gap function. Both equations acquire modifications due to interplay of disorder and interactions at scales shorter than superconducting coherence length. We solve these equations for the cases of Ising-type and strong spin-orbit couplings. In the former case one triplet diffusive mode remains effective at long lengthscales whereas in the latter case all triplet modes are suppressed. In both cases we determine superconducting transition temperature and determine the energy dependence of the spectral gap function at low temperatures, $T\ll T_c$, and in the vicinity of the transition, $T_c-T\ll T_c$. We find that the maximal magnitude of the gap is proportional to $T_c$, i.e. it is also enhanced by multifractality. In addition, we estimate the mesoscopic fluctuations of the local density of states. We find that these fluctuations are logarithmically divergent with the system size (if one neglects the effect of dephasing) in spite of the absence of spin rotational symmetry.

The outline of the paper is as follows. In Sec. \ref{Sec:SigmaModel} we present the general scheme for description of the superconducting state. This scheme is applied to the case of Ising-type spin-orbit coupling in Sec. \ref{Sec:IsingSOC}. In Sec. \ref{Sec:Interaction} we consider the case of strong spin-orbit coupling. The fluctuations of the local density of states are studied in Sec. \ref{Sec:LDOS}. We end the paper with summary and conclusions in Sec. \ref{Sec:DiscConc}. Some technical details are delegated to Appendices.

\section{Equation for the spectral gap 
\label{Sec:SigmaModel}}

In dirty superconductors there is substantial energy window between diffusive scale $1/\tau$ ($\tau$ stands for the mean free time) and the energy scale related with superconductivity for which one can choose $T_c$. Therefore, in order to describe superconducting properties which correspond typically to the energy scale $T_c$ (the corresponding length scale is the superconducting coherence length, $\xi=\sqrt{D/T_c}$) 
one has to take into account effects related with interaction of diffusive modes in the energy interval $T_c\lesssim \varepsilon \lesssim 1/\tau$. As well-known from studies of normal dirty metals, the main effect of diffusive modes is renormalization of physical parameters of the system, e.g. conductance, interaction strengths, etc. 

Such renormalization should be taken into account for the superconducting state as well. The most profound effect of the renormalization is the modification of the Usadel equation and the self-consistency equation for the spectral gap. These modified equations can be derived by means of the nonlinear sigma model following approach of Ref. \cite{BurmistrovAnnPhys2021} (see details in Appendix A). This procedure results in the following modified Usadel equation for the spectral angle $\theta_\varepsilon$,
\begin{gather}
 \frac{D_\varepsilon}{2}\nabla^2\theta_\varepsilon  -|\varepsilon| \sin \theta_\varepsilon + \Delta_\varepsilon \cos\theta_\varepsilon = 0 . \label{eq:NLSM:Usadel}
\end{gather}
Here $\varepsilon=\pi T(2n+1)$ denotes fermionic Matsubara frequency.
Equation \eqref{eq:NLSM:Usadel} differs from the standard Usadel equation \cite{Usadel} by energy dependent spectral gap $\Delta_\varepsilon$ and energy dependent diffusion coefficient $D_\varepsilon$ \footnote{In this paper we are interested in the superconducting state which is spatially homogeneous on the scale of the order of $\xi$. Therefore, we shall not discuss energy dependence of the diffusion coefficient here.}. 

To the lowest order in disorder and interaction, the spectral gap satisfies the following equation
\begin{align} 
\Delta_\varepsilon  = &  -2\pi T\sum_{\varepsilon^\prime_n>0} \sin\theta_{\varepsilon^\prime} \bigg\{\gamma_c  - 2\frac{(\gamma_s - \mathcal{N} \gamma_t)}{g}  \notag \\
   \times & \int \frac{d^2 \bm{q}}{(2\pi)^2}
   \frac{D}{D q^2+E_\varepsilon+E_{\varepsilon^{\prime}}}
   \bigg\} ,\notag \\
   & E_\varepsilon = |\varepsilon| \cos \theta_\varepsilon + \Delta \sin \theta_\varepsilon .
   \label{eq:Delta:eps}
\end{align}

\noindent  Here $\gamma_c<0$, $\gamma_s$, and $\gamma_t$ stands for the bare values of dimensionless interaction amplitudes in the Cooper channel as well as in the singlet and triplet particle-hole channels, respectively. We assume the interaction in the particle-hole channel to be weak and short-ranged. Therefore, we consider the case of $|\gamma_{c,s,t}|\ll 1$.

The disorder is controlled by the bare dimensionless (in units $e^2/h$) conductance $g=h/(e^2 R_\square)$ where $R_\square$ is the film resistance per square in the normal state. The bare diffusion coefficient $D$ is related with conductance and the density of states, $\nu$, at the Fermi energy in the normal state by means of the Einstein relation $g=2\pi \nu D$. 
The superconducting order parameter $\Delta$ determines the bare value of the superconducting gap.

Parameter $\mathcal{N}$ in Eq. \eqref{eq:Delta:eps} counts the number of massless triplet diffusive modes. We shall concentrate below on the cases $\mathcal{N}=0$ and $1$ whereas the case $\mathcal{N}=3$ was considered in Ref. \cite{BurmistrovAnnPhys2021}. 

We note that a similar form of the self-consistency equation for the spectral gap has been derived in Ref. \cite{Skvortsov2005,Skvortsov2012} in the case of Coulomb interaction ($\gamma_s=-1$) and neglect of exchange interaction ($\gamma_t=0$) by means of the diagrammatic technique. 

Equation \eqref{eq:Delta:eps} resembles the standard self-consistency equation in the BCS theory except logarithmic renormalization of the attraction interaction parameter $\gamma_c$. 
This renormalization is exactly the same as in the normal metal except the infrared scale is set by $\max\{\varepsilon,\varepsilon^\prime,\Delta\}$. The perturbative result \eqref{eq:Delta:eps} for the renormalization of $\gamma_c$
can be extended by means of the renormalization group technique, see Ref. \cite{BurmistrovAnnPhys2021} for details. 

Then solving the modified Usadel equation \eqref{eq:NLSM:Usadel} as $\sin\theta_\varepsilon =\Delta_\varepsilon/\sqrt{\varepsilon^2+\Delta_\varepsilon^2}$, we find the following self-consistency relation for $\Delta_\varepsilon$:
\begin{equation}
    \Delta_\varepsilon = -2\pi T\sum_{\varepsilon^\prime>0} \frac{ \gamma_c\left(L_{E_{\varepsilon}+E_{\varepsilon^\prime}}\right) \Delta_{\varepsilon^\prime}}{\sqrt{\varepsilon^{\prime 2}+\Delta_{\varepsilon^\prime}^2}}, \label{eq:NSLM:sc-general}
\end{equation}
Here $L_\varepsilon = \sqrt{D/\varepsilon}$ is the diffusive length associated with the energy scale $\varepsilon$. The flow of $\gamma_c$ with the length scale $L$ is governed by the following renormalization group equation (see Ref. \cite{BurmistrovAnnPhys2021} for $\mathcal{N}=3$),
\begin{equation}
    \frac{d\gamma_c}{dy} = - \frac{t}{2}(\gamma_s - \mathcal{N} \gamma_t).
    \label{eq:RG:gamma:c}
\end{equation}

\noindent Here $y = \ln L/\ell$ with $\ell$ and $L$ being the mean free path and the system size, respectively. The dimensionless resistance is denoted as $t = 2/(\pi g)$. Its bare value $t_0$ is assumed to be small, $t_0\ll 1$. 

Equation \eqref{eq:RG:gamma:c} does not contain the standard term, -$\gamma_c^2$, which is responsible for the Cooper instability in the clean case. This term is encoded in the superconducting order parameter $\Delta$ (see Ref. \cite{BurmistrovAnnPhys2021} for details).

It is tempting to substitute $\Delta$ by $\Delta_\varepsilon$ in the expression for $E_\varepsilon$ in Eq. \eqref{eq:NSLM:sc-general} to make it fully self-consistent equation for $\Delta_\varepsilon$. This is exactly what was done in Ref. \cite{BurmistrovAnnPhys2021} based on the relation between the Usadel equation linearized in variation of $\theta_\varepsilon$ and the Cooperon propagator with coinciding energies. However, further analysis demonstrates that Cooperon propagator with two non-equal Matsubare energies has more complicated structure after renormalization. \footnote{The authors are grateful to P. Nosov for this comment.} Fortunately, as we shall see below the precise form of difference $E_\varepsilon -|\varepsilon|$ is not essential for results reported in this paper.    

Eq. \eqref{eq:RG:gamma:c} is needed to  be supplemented by 
renormalization group equations for $\gamma_{s,t}$ and $t$. However, their precise form depends on the magnitude of $\mathcal{N}$. Below we shall analyse Eq. \eqref{eq:NSLM:sc-general} separately for the cases $\mathcal{N}=0$ and $1$. In what follows we assume that bare values of interaction and disorder are weak: $|\gamma_{s0}|, |\gamma_{t0}|, |\gamma_{c0}|, t_0 \ll 1$.

\section{$\mathcal{N}=1$: Ising spin-orbit coupling \label{Sec:IsingSOC}}

In this section we focus on superconducting films with the so-called Ising spin orbit coupling. The latter pins electron spins to the out-of-plane direction. In this case in-plane spin-flip scattering rates, induced by spin-orbit coupling are smaller than the out-of-plane one,  $1/\tau_{\text{so}}^{x,y}\ll 1/\tau_{\text{so}}^{z}$. Hence, one triplet diffusive mode, corresponding to the total spin projection $S_z=0$, remains gapless as well as singlet
diffusive mode. 
That is why the case $\mathcal{N}=1$ is realized for the Ising spin-orbit coupling.


In order to analyse the gap equation \eqref{eq:NSLM:sc-general} one needs to know the actual dependence of $\gamma_c$ on the length scale. In Ref. \cite{BurmistrovPRB2015} the full set of one-loop renormalization group equations for $\gamma_{s,t,c}$ and $t$ has been derived by means of the background field  renormalization of the nonlinear sigma model above superconducting transition temperature. Applying the same method in the superconducting state, we find
\begin{subequations}
\begin{align}
    \frac{dt}{dy} &= -\frac{t^2}{2} (\gamma_s + \gamma_t + 2\gamma_c),  \label{eq:N=1:t-RG} 
     \\
    \frac{d}{dy}\begin{pmatrix} \gamma_s \\ \gamma_t \\ \gamma_c \end{pmatrix} & = - \frac{t}{2}
    \begin{pmatrix}
    1 & 1 & 2 \\
    1 & 1 & -2 \\
    1 & -1 & 0
    \end{pmatrix}
    \begin{pmatrix} \gamma_s \\ \gamma_t \\ \gamma_c \end{pmatrix} .
    \label{eq:N=1:RG}
    \end{align}
\end{subequations}
We note that Eq. \eqref{eq:N=1:t-RG} is written under assumption that $|\gamma_{s,t,c}|\ll 1$. In the same way as in the normal metal  \cite{Lee},   weak localization and weak antilocalization compensate each other in the presence of Ising spin-orbit coupling.
Eq. \eqref{eq:N=1:t-RG} implies that dimensionless resistance $t$ remains almost constant in the leading order. 
Therefore, we shall assume $t\simeq t_0$ below. Equation \eqref{eq:N=1:RG} suggests that under the renormalization group flow the interaction parameters approach the so-called the BCS line $-\gamma_s = \gamma_t = \gamma_c \equiv \gamma$ \cite{BurmistrovPRL2012}. In order to describe the behavior of the system at  lengthscales $y\gtrsim t_0^{-1}$, we project Eq. \eqref{eq:N=1:RG} onto the BCS line. Then we shall work with effective interaction parameter $\gamma$ that flows according to
\begin{gather}
    d\gamma/dy \simeq t_0 \gamma, \;\;\; \gamma_0 = (\gamma_{t0}-\gamma_{s0}+2\gamma_{c0})/4 < 0.
\end{gather}
 Solving the above equation, we find 
\begin{equation}
\gamma(L) = \gamma_0 (L/\ell)^{t_0} . \label{eq:RG:gamma_0}
\end{equation}


Below we assume that disorder dominates interaction, i.e. $t_0\gg|\gamma_0|$. As known \cite{BurmistrovPRL2012}, it is the regime in which the multifracal enhancement of superconducting transition temperature is expected.
The transition temperature $T_c$ can be estimated from the relation $|\gamma(L_{T_c})|\sim t_0$ (see Appendix B). It yields 
\begin{equation}
T_c\sim (1/\tau) (|\gamma_0|/t_0)^{2/t_0} . \label{eq:N=1:Tc-RG}
\end{equation}
We note that the superconducting critical temperature $T_c$ corresponds to $y_c \sim t_0^{-1} \ln (t_0/|\gamma_0|) \gg t_0^{-1}$. This justifies the projection onto the BCS line.

\subsection{The critical temperature \label{SubSec:Tc}}

More accurately the superconducting transition temperature $T_c$ can be determined from the linearized self-consistent equation, cf. Eq. \eqref{eq:NSLM:sc-general}. We note that after projection of the self-consistency equation \eqref{eq:NSLM:sc-general} onto the BCS line, $\gamma_c$ in it is substituted by $\gamma$. Linearized version of the self-consistency equation then reads
\begin{equation}
    \Delta_\varepsilon = -2\pi T\sum_{\varepsilon^\prime>0} \gamma(L_{\varepsilon+\varepsilon^\prime}) \frac{\Delta_{\varepsilon^\prime}}{\varepsilon^\prime} .
    \label{eq:N=1:sc-linearized}
\end{equation}

\noindent{Taking} into account the actual dependence of $\gamma(L)$ on $L$, Eq. \eqref{eq:RG:gamma_0}, we find
\begin{gather}
    \Delta_n = \frac{|\gamma_0|}{(2\pi T \tau)^{t_0/2}} \sum_{n' \geqslant 0}^{n_{\rm max}} \frac{\Delta_{n'}}{(n+n'+1)^{t_0/2}(n'+1/2)} , \label{eq:N=1:sc-Tc}
\end{gather}
where $n_{\rm max}\simeq 1/(2\pi T\tau)$ is the natural cut off for number of Matsubara frequencies belonging to the diffusive regime. The search for $T_c$ in Eq. \eqref{eq:N=1:sc-Tc} can be reformulated as a problem of the maximal eigenvalue of the corresponding matrix. The superconducting transition temperature satisfies 
$\left(2\pi T_c \tau\right)^{t_0/2}=|\gamma_0|\lambda_{M}$ where $\lambda_{M}$ is the maximal eigenvalue of the matrix $M_{nn'} = (n+n'+1)^{-t_0/2}(n'+1/2)^{-1}$. 

\begin{figure}[t]
\centerline{
    \includegraphics[width=0.46\textwidth]{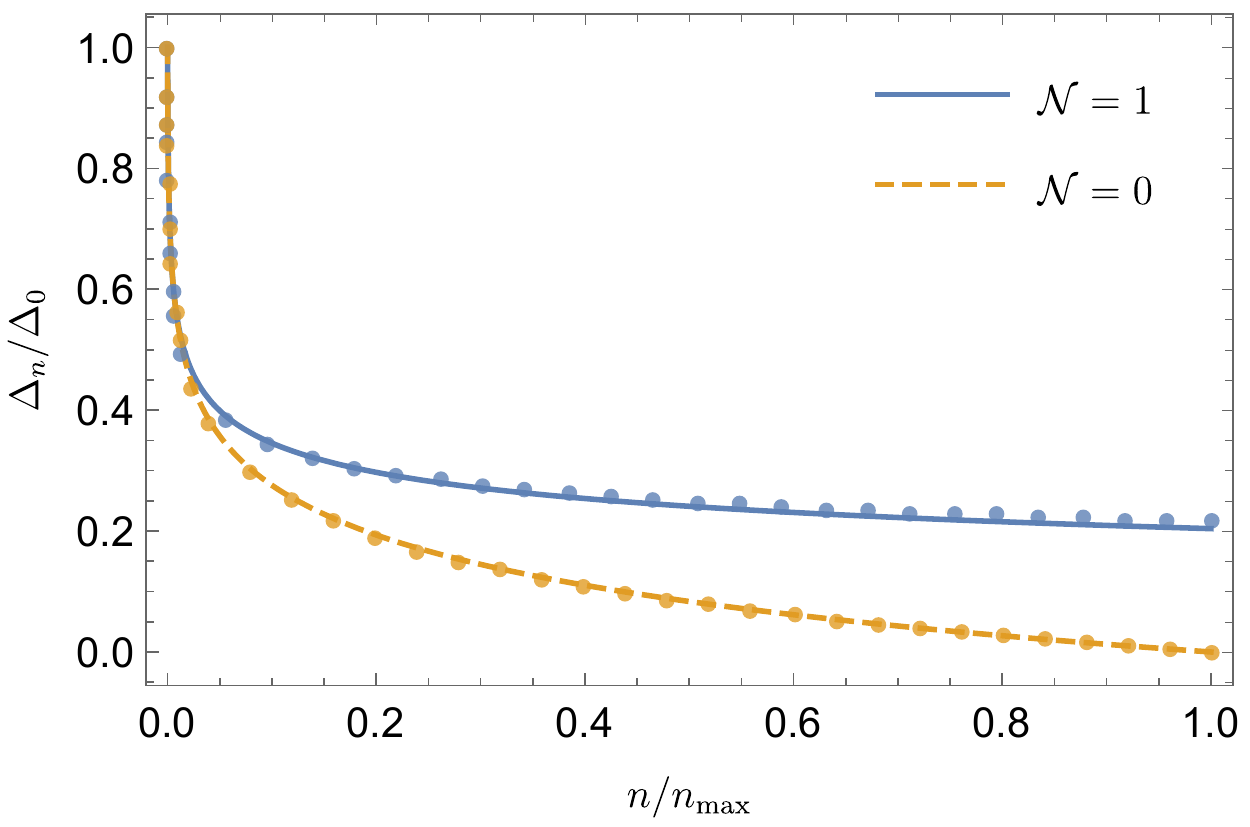}}
     \caption{The dependence of the gap function $\Delta_n$ on the Matsubara energies $\varepsilon_n = 2\pi T(n+1/2)$ for $T$ near the critical temperature $T_c$. Solid and dashed lines show analytical expressions \eqref{eq:N=1:f(u)} and \eqref{eq:N=0:f(u)}, respectively. Dots of the corresponding colors mark numerical solutions for leading eigenvectors of Eqs. \eqref{eq:N=1:sc-Tc} and \eqref{eq:N=0:sc-Tc}. 
     }
      \label{fig-Delta-near-Tc}
\end{figure}

Numerical solution of Eq. \eqref{eq:N=1:sc-Tc} by the means of the power method (see Appendix C) gives
$\lambda_{M}\simeq 1.4/t_0$ such that
\begin{equation}
    T_c \simeq  \frac{1}{2\pi \tau}  (1.4 |\gamma_0|/t_0)^{2/t_0}.  \label{eq:N=1:Tc-num} 
\end{equation}
The right eigenvector $r_n$ of the matrix $M$ corresponding to $\lambda_{M}$ is shown in Fig.~\ref{fig-Delta-near-Tc}. (The left eigenvector of the matrix $M$ is expressed  as $l_n=r_n/(n+1/2)$.) There is a significant energy dependence of the spectral gap in contrast with the BCS model for which 
it is just a constant.

The result \eqref{eq:N=1:Tc-num} can be also justified on the basis of 
analytical treatment of Eq. \eqref{eq:N=1:sc-Tc}. Let us first replace $(n+n^\prime+1)^{t_0/2}$ with $\max\{(n+1/2)^{t_0/2}, (n^\prime+1/2)^{t_0/2}\}$. It is justified by the smallness of exponent $t_0 \ll 1$. Next we introduce a variable 
\begin{equation}
u_\varepsilon = \frac{2}{t_0} |\gamma(L_\varepsilon)| = \frac{2|\gamma_0|}{t_0}(\varepsilon \tau)^{-t_0/2} .
\label{eq:ueps:def}
\end{equation}
Then, employing the Euler-Maclaurin resummation to the right hand side of Eq. \eqref{eq:N=1:sc-linearized}, we obtain
\begin{subequations}
\begin{gather}
    \Delta_{u_n} \simeq u_n \int \limits_{u_{n}}^{u_{0}} du \frac{\Delta_u}{u} +  \int \limits_{u_{\infty}}^{u_{n}} du \Delta_u + \frac{a t_0}{2} u_n \Delta_{u_0} , \label{eq:N=1:sc-EulerMaclaurin}\\
    u_\infty\equiv u_{1/\tau}\sim |\gamma_0|/t_0 \ll 1 ,  \\
   a=1+\sum_{k=1}^\infty 2^{2k-1} B_{2k}/k \approx 1.27 . \label{eq:N=1:a-def}
\end{gather}
\end{subequations}

\noindent Here $u_\infty$ corresponds to $n_{\max}$ and  
 $B_{2k}$ denotes even Bernoulli numbers. 
 At $T=T_c$ we seek the solution of Eq. \eqref{eq:N=1:sc-EulerMaclaurin} in the form $\Delta_{u_n} = \Delta_{u_0} f(u_n)$ with the normalization $f(u_0)=1$. The integral equation \eqref{eq:N=1:sc-EulerMaclaurin} can be reduced to the following Cauchy problem for the unknown function $f(u)$, 
\begin{gather}
    f^{\prime \prime}(u) = -f(u)/u, \notag\\
    f^\prime(u_0) = a t_0/2, \qquad f^{\prime}(u_\infty) = f(u_\infty)/u_\infty . \label{eq:N=1:f(u)-boundary}
\end{gather}

\noindent Solving Eq. \eqref{eq:N=1:f(u)-boundary}, we obtain
\begin{gather}
    f(u) = \frac{F_1(u)}{F_1(u_0)}, \;\;\; F_1(u) \simeq \sqrt{u} J_1(2\sqrt{u}). \label{eq:N=1:f(u)} 
\end{gather}

\begin{figure}[t]
\centerline{
    \includegraphics[width=0.5\textwidth]{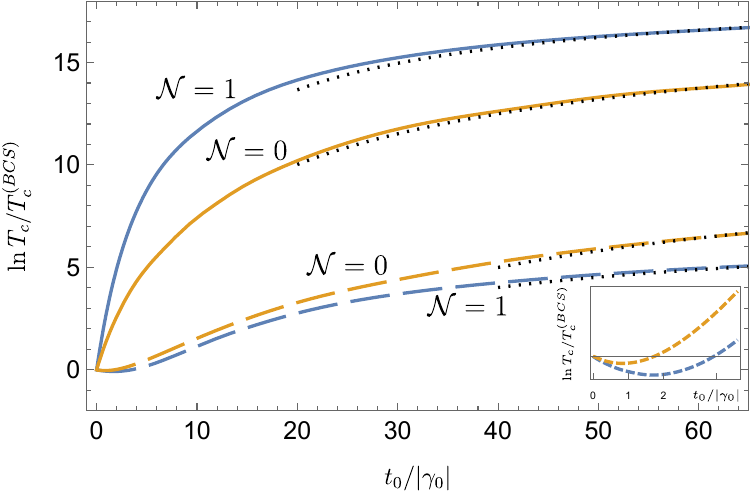}
    }
     \caption{Multifractally-increased superconducting transition temperature $T_c$. Solid lines show the dependence of $\ln T_c/T_c^{(BCS)}$ on magnitude of the ratio $t_0/|\gamma_0|$, when the initial interaction parameters $\gamma_{s0}$, $\gamma_{t0}$, $\gamma_{c0}$ lie on the BCS-line, i.e. $-\gamma_s = \gamma_t = \gamma_c = \gamma$ for $\mathcal{N} = 1$ and $-\gamma_s = \gamma_c = \gamma$ for $\mathcal{N} = 0$. Dashed lines of the corresponding colors illustrate the behavior of $\ln T_c/T_c^{(BCS)}$, when the initial parameters deviate from the BCS-line. This can lead to a decrease in the critical temperature $T_c$ in comparison to $T_c^{(BCS)}$ at a small ratio of $t_0$ to $|\gamma_0|$ (see inset), but eventually $\ln T_c/T_c^{(BCS)}$ becomes positive and continues to grow with increasing disorder. Black dotted lines correspond to expressions \eqref{eq:N=1:Tc-EulerMaclaurin} and \eqref{eq:N=0:Tc-EulerMaclaurin}.
     }
      \label{figTc}
\end{figure}

\noindent Here $J_1(x)$ denotes the Bessel function of the first kind. We note that for the sake of brevity in the above expression $F_1(u)$ is written in the lowest order in small parameters $|\gamma_0| \ll t_0 \ll1$. Although one can easily find the exact solution to $f(u)$, in what follows we do not need it. We also note that the solution \eqref{eq:N=1:f(u)} satisfies the normalization condition $f(u_0)=1$ and the boundary condition at $u=u_\infty$. The yet unknown parameter $u_0$ determines the superconducting transition temperature as
\begin{equation}
T_c = (2\pi\tau)^{-1} ((2/u_0) |\gamma_0|/t_0)^{2/t_0}. 
\label{eq:N=1:Tc-EulerMaclaurin}
\end{equation}
It can be extracted from the boundary condition at $u=u_0$. Using the relation $(x J_1(x))^\prime=x J_0(x)$ and neglecting $a t_0/2$ in the right hand side of the equation for the boundary condition at $u=u_0$, we find $u_0 \simeq (j_{0,1})^2/4 \approx 1.45$ where $j_{n,k}$ is the $k$-th zero of the Bessel $J_n(x)$ function. The result \eqref{eq:N=1:Tc-EulerMaclaurin} is in quantitative agreement with the result \eqref{eq:N=1:Tc-num} of numerical approach. Moreover, as shown in Fig. \ref{fig-Delta-near-Tc}, there is a remarkable agreement between the function $f(u_n)$ and numerically computed eigenvector corresponding to the maximal eigenvalue of the matrix $M$.

Equation \eqref{eq:N=1:Tc-num} predicts increase of $T_c$ with increase of disorder, $t_0$, at fixed $\gamma_0$. However, Eq. \eqref{eq:N=1:Tc-num} is valid for $t_0\gg|\gamma_0|$ only. We solve the self-consistency equation \eqref{eq:N=1:sc-Tc} for various $t_0$ numerically. The obtained dependence of $T_c$ on $t_0$ is shown in Fig. \ref{figTc}. As one can see, for initial conditions at the BCS line $T_c$ grows with increase of $t_0$ reaching  
the asymptotic expression \eqref{eq:N=1:Tc-EulerMaclaurin} (black dotted lines) at $t_0\gg|\gamma_0|$. In case of the system away from the BCS line initially, at first, $T_c$ is suppressed with increase of $t_0$, but then it starts to grow at $t_0\gtrsim|\gamma_0|$.


We now move on to 
examining the behaviour of the gap function $\Delta_\varepsilon$ as a function of $\varepsilon$ at different temperature regimes: when $T$ is close to $T_c$ and when $T \ll T_c$.


\subsection{The spectral gap function \label{SubSec:GapFunction}}

\subsubsection{The gap function near $T_c$}

At $T = T_c$ the amplitude $\Delta_0$ of the spectral gap function vanishes. To find the dependence of $\Delta_\varepsilon$ on energy at $T_c-T\ll T_c$, we expand the modified self-consistency equation to the third order
, retrieving
\begin{equation}
    \Delta_\varepsilon = 2\pi T \sum_{\varepsilon^\prime>0} |\gamma(L_{\varepsilon+\varepsilon^\prime})| \left( \frac{\Delta_{\varepsilon^\prime}}{\varepsilon^\prime} - \frac{\Delta_{\varepsilon^\prime}^3}{2\varepsilon^{\prime 3}} \right). \label{eq:N=1:sc-nearTc}
\end{equation}
We note that quadratic in $\Delta_\varepsilon$ terms that originate from expansion of $\big|\gamma(L_{E_{\varepsilon}+E_{\varepsilon^{\prime}}})\big|$ are suppressed by a small factor $t_0 \ll 1$.

Let us write $\Delta_{\varepsilon_n}=\Delta_0(T) r_n$ with the normalization $r_0=1$. Then Eq. \eqref{eq:N=1:sc-nearTc} becomes
\begin{gather}
    \lambda_M \left(\frac{T}{T_c}\right)^{\frac{t_0}{2}} r_n = \sum_{n^\prime=0}^{n_{\max}} M_{nn^\prime} \left[r_{n^\prime} - \frac{\frac{\Delta_0^2(T)}{8\pi^2 T^2_c} r_{n^\prime}^3}{(n^\prime+1/2)^2}  \right] . \label{eq:N=1:sc-Tc-rn}
\end{gather}
In order $\Delta_0(T_c)=0$, one need to choose $r_n$ to be the right eigenvector of the matrix $M_{nn^\prime}$, corresponding to its maximal eigenvalue $\lambda_M$. Then, multiplying both sides of Eq. \eqref{eq:N=1:sc-Tc-rn} on the left eigenvector $l_n=r_n/(n+1/2)$ (corresponding to $\lambda_M$) from the left, we find
\begin{equation}
    \Delta_0(T) = \left (b_{\mathcal{N}} \frac{8\pi^2}{7\zeta(3)}  T_c(T_c-T) \right)^{1/2} . \label{eq:N=1:Tc-Delta0(T)} 
\end{equation}    
 Here the constant $b_{\mathcal{N}}$ for  $\mathcal{N}=1$ is given as   
\begin{equation}    
    b_1 = \frac{7\zeta(3) t_0}{2} 
\frac{\sum\limits_{n=0}^{n_{\max}} r^2_n/(n+1/2)}{\sum\limits_{n=0}^{n_{\max}}r_n^4/(n+1/2)^3} . 
    \label{eq:N=1:Tc-b}
\end{equation}
We chose normalization of $b_1$ in such a way that the quantity $b_1-1$ describes the deviation from the BCS theory. Although, one could expect the nonzero value for the difference  $b_1-1$ due to strong energy dependence of the right eigenvector, as we shall demonstrate below, this is not the case within our approximation.

First of all, replacing $r_n^4$ with $r_0^4 = 1$ in the denominator of Eq. \eqref{eq:N=1:Tc-b}, we find that 
\begin{equation}
    \sum\limits_{n=0}^{n_{\max}} \frac{r_n^4}{(n+1/2)^3} \simeq 7 \zeta(3). 
\end{equation}
Next, in order to estimate the numerator of Eq. \eqref{eq:N=1:Tc-b}, we replace $r_n$ with an analytical expression $r_n = f(u_n)$, see Eq. \eqref{eq:N=1:f(u)}. Hence, we obtain
\begin{gather}
    \sum\limits_{n=0}^{n_{\max}} \frac{r_n^2}{n+1/2} \simeq \frac{2}{t_0} \int_{0}^{u_0} \frac{du}{u} f^2(u) \simeq \frac{2}{t_0}. 
\end{gather}
Combining all the above we restore the BCS results, $b_1 = 1$, in the limit $t_0\ll 1$.
We note that in order to compute corrections to the BSC result, $b_1=1$, one needs to know the precise form of dependence of the infrared cut off length scale $L_{E_{\varepsilon}+E_{\varepsilon^\prime}}$ on $\Delta_\varepsilon$ and $\Delta_{\varepsilon^\prime}$ in Eq. \eqref{eq:NSLM:sc-general}.

One can estimate the effect of admixture of other eigenmodes to the dependence of $\Delta_\varepsilon$ on $\varepsilon$. Writing $\Delta_\varepsilon=\Delta_0(T) (r_n+\sum_j s_j r_n^{(j)})$ where $r_n^{(j)}$ are the right eigenvectors of the matrix $M$ with eigenvalues $\lambda_j<\lambda_M$, we find
\begin{gather}
    s_j = -\frac{\lambda_j}{\lambda_M-\lambda_j}\frac{\Delta_0^2(T)}{8\pi^2 T_c^2} \frac{\sum\limits_{n=0}^N l_n^{(j)} r_n^3/(n+1/2)^2}{\sum\limits_{n=0}^N l_n^{(j)} r_n^{(j)}}  .
\end{gather}
Here we used orthogonality condition $\sum_n l_n^{(j)} r_n =0$ where $l_n^{(j)}$ stands for the left eigenvector of the matrix $M_{nn^\prime}$ corresponding to the eigenvalue $\lambda_j$. We see that the admixture of the other eigenmodes at $T_c-T\ll T_c$ is completely negligible. Therefore, the energy dependence of $\Delta_\varepsilon$ on $\varepsilon$ at $T_c-T\ll T_c$ is essentially the same as at the transition.

\subsubsection{The gap function at 
$T \ll T_c$\label{Sec:gapT0:N1}}

We start from the zero-temperature limit. To study the behaviour of the gap function at $T=0$ we substitute summation over Matsubara frequencies by integration over energy $\varepsilon^\prime$ in Eq. \eqref{eq:NSLM:sc-general}. We note that there are two sources for dependence on $\varepsilon^\prime$. The fast dependence under the square root and slow (almost logarithmic for $t_0\ll 1$) dependence in $\gamma$. On the basis of the solution for $T=T_c$ we expect $\Delta_\varepsilon$ to be decreasing function of $\varepsilon$. 
Let us introduce the characteristic energy scale $\varepsilon_0$, such that $\Delta_{\varepsilon_0} = \varepsilon_0$. The structure of the equation on $\Delta_\varepsilon$, see Eq. \eqref{eq:NSLM:sc-general}, suggests that the gap-function slightly varies from its value $\Delta_0$ at $\varepsilon = 0$ to 
$\Delta_{\varepsilon_0}=\varepsilon_0$ at $\varepsilon = \varepsilon_0$. This implies that $\varepsilon_0\sim \Delta_0$. In order to find the precise relation between $\varepsilon_0$ and $\Delta_0$ as well as dependence of $\Delta_\varepsilon$ at $\varepsilon<\Delta_0$, one needs to know the precise dependence of the infrared length scale $L_{E_{\varepsilon}+E_{\varepsilon^\prime}}$ on $\Delta_\varepsilon$ and $\Delta_{\varepsilon^\prime}$, see Eq. \eqref{eq:NSLM:sc-general}. 
This complication is absent at large energies $\varepsilon\gg\Delta_\varepsilon$. As we shall check below, the latter condition is fulfilled at energies $\varepsilon$ not too close to $\Delta_0$, i.e. at $\varepsilon\gtrsim\Delta_0$, due to the smallness of dimensional resistance, $t_0\ll 1$.  

At $\varepsilon \gg \Delta_\varepsilon$ we approximate the self-consistent equation \eqref{eq:NSLM:sc-general} as follows
\begin{gather}
    \nonumber \Delta_\varepsilon \simeq |\gamma(L_\varepsilon)| \int\limits_0^{\Delta_0} \frac{d \varepsilon^\prime \Delta_0}{\sqrt{\varepsilon^{\prime 2} + \Delta^2_0}} + |\gamma(L_\varepsilon)| \int\limits_{\Delta_0}^\varepsilon \frac{d\varepsilon^\prime \Delta_{\varepsilon^\prime}}{\sqrt{\varepsilon^{\prime 2}+\Delta^2_{\varepsilon^\prime}}} \\
    + \int\limits_\varepsilon^{1/\tau} \frac{d \varepsilon^\prime \Delta_{\varepsilon^\prime}}{\sqrt{\varepsilon^{\prime 2}+\Delta^2_{\varepsilon^\prime}}} |\gamma(L_{\varepsilon^\prime})|. \label{eq:N=1:T=0:e>e0}
\end{gather}
Substituting $\Delta_\varepsilon$ for $\Delta_{u_\varepsilon} = \Delta_{u_0} f(u_\varepsilon)$, where $u_\varepsilon$ is defined in Eq. \eqref{eq:ueps:def}, the above equation can be rewritten in the following differential form
\begin{equation}
f^{\prime\prime}(u_\varepsilon) =
-  \frac{\varepsilon f(u_\varepsilon)/u_\varepsilon}{\sqrt{\varepsilon^2+\Delta_0^2f^2(u_\varepsilon)}} .
\label{eq:eqL12}
\end{equation}
Since we are working in the regime $\varepsilon\gg \Delta_\varepsilon$, we can safely neglect the term with $\Delta_0$ under the square root in the right hand side of Eq. \eqref{eq:eqL12}. Then Eq. \eqref{eq:eqL12} reduces to Eq. \eqref{eq:N=1:f(u)-boundary} with the same boundary conditions except $a=\arcsinh(1)$ now. Thus the solution can be read from 
Eq. \eqref{eq:N=1:f(u)}. Altogether we find the following solution for the spectral gap function
\begin{gather}
    \Delta_\varepsilon \simeq \Delta_0 
    \begin{cases}
        1, & \varepsilon \lesssim \Delta_0, \\
        F_1(u_\varepsilon)/F_1(u_{\Delta_0}), & \varepsilon \gtrsim \Delta_0 .
    \end{cases} 
    \label{eq:Delta:eps:T0}
\end{gather} 
Now we can check the assumption $\varepsilon\gg\Delta_\varepsilon$. Using Eq. \eqref{eq:Delta:eps:T0}, we find that
\begin{equation}
\frac{\varepsilon}{\Delta_\varepsilon}=
\frac{u_{\Delta_0}^{2/t_0}F_1(u_{\Delta_0})}{u_\varepsilon^{2/t_0}F_1(u_\varepsilon)}
\gg 1
\end{equation}
holds for all $\varepsilon$ except close vicinity of $|\varepsilon-\Delta_0|\sim\Delta_0$.

We note perfect matching of both asymptotic results \eqref{eq:Delta:eps:T0} at $\varepsilon=\Delta_0$. Using the boundary condition $f^\prime(u_{\Delta_0})=0$ we find that $\Delta_0$ coincides with $T_c$ upto numerical prefactor,
\begin{equation}
    \Delta_0\sim T_c .
    \label{eq:Delta_0:exp}
\end{equation}
However, as we noted above our approach does not applicable in the vicinity of the point $\varepsilon=\Delta_0$. Therefore, we cannot determine the precise constant for the ratio for $\Delta_0/T_c$. We emphasize that the dependence of the spectral gap function at $\varepsilon\gg \Delta_0$ for $T=0$ is exactly the same as for $T$ close to $T_c$.  

\begin{figure}[t]
\centerline{
    \includegraphics[width=0.46\textwidth]{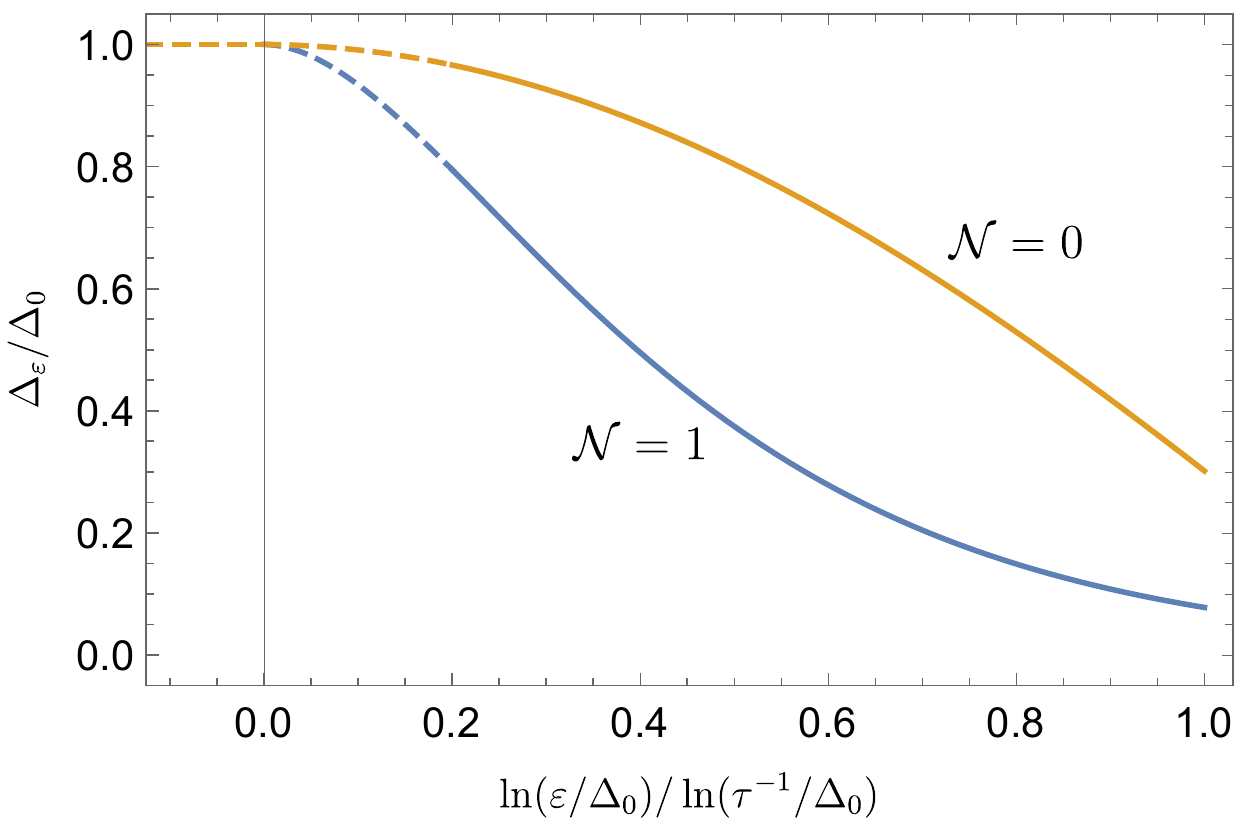}
    }
     \caption{
     The dependence of the gap function $\Delta_\varepsilon$ at low temperatures on energy $\varepsilon$ (see text). Dimensionless interaction constant is chosen to be $|\gamma_0| = 0.005$ and dimensionless resistance is $t_0 = 0.2$. Bottom curve illustrates the behaviour of $\Delta_\varepsilon$ for the case of Ising spin-orbit coupling ($\mathcal{N}=1$) while the top one corresponds to the case of strong spin-orbit coupling  ($\mathcal{N}=0$).}
      \label{fig1}
\end{figure}

We emphasize that the spectral gap function is parametrically enhanced at small energies. Using Eq. \eqref{eq:Delta:eps:T0} we find that the spectral gap at $\varepsilon\sim 1/\tau$ (which coincides with the order parameter $\Delta$) is proportional to $\Delta_0 |\gamma_0|/t_0 \ll \Delta_0$. Typical behaviour of $\Delta_\varepsilon$ is illustrated in Fig. \ref{fig1}.

At non-zero temperature, the form of the spectral gap function remains the same but there is reduction of the magnitude of $\Delta_0$ while temperature increases. At $T\ll\Delta_0$ the dependence of $\Delta_0$ on temperature can be estimated as follows. At $\varepsilon\gg\Delta_0$ the gap function $\Delta_\varepsilon$ satisfies equation similar to Eq. \eqref{eq:N=1:T=0:e>e0} in which $\Delta_0$ is substituted by $\Delta_0(T)$ and
\begin{gather}
\int\limits_0^{\Delta_0} \frac{d \varepsilon^\prime \Delta_0}{\sqrt{\varepsilon^{\prime 2} + \Delta^2_0}}    \to 2\pi T\sum_{\varepsilon^\prime>0}\frac{\Delta_0(T)}{\sqrt{\varepsilon^{\prime 2} + \Delta^2_0(T)}}
\notag\\
-\int\limits_{\Delta_0(T)}^{1/\tau}\frac{d \varepsilon^\prime \Delta_0(T)}{\sqrt{\varepsilon^{\prime 2} + \Delta^2_0(T)}} \simeq \int\limits_0^{\Delta_0(T)} \frac{d \varepsilon^\prime \Delta_0(T)}{\sqrt{\varepsilon^{\prime 2} + \Delta^2_0(T)}} 
\notag \\
- \sqrt{\frac{2\pi T}{\Delta_0}} e^{-\Delta_0/T} .
\end{gather}
Such modification of Eq. \eqref{eq:N=1:T=0:e>e0} results in change of the constant $a$ in the boundary conditions in Eq. \eqref{eq:N=1:f(u)-boundary}. Now it becomes $a=\arcsinh(1)-\sqrt{2\pi T/\Delta_0} \exp(-\Delta_0/T)$. Taking this temperature shift of the constant $a$ into account, we find 
\begin{equation}
   \Delta_0 - \Delta_0(T) \sim \sqrt{2\pi T \Delta_0} \; e^{-\Delta_0/T}, \quad T\ll\Delta_0 .
   \label{eq:Delta:T}
\end{equation}
We note that since this result is obtained with the help of the boundary condition at $\varepsilon=\Delta_0(T)$, we cannot unambiguously  determine a numerical factor in Eq. \eqref{eq:Delta:T}.

\section{$\mathcal{N}=0$: Strong spin-orbit coupling \label{Sec:Interaction}}

In this section we consider the case when all three  spin-flip rates, $1/\tau_{\text{so}}^{x,y,z}$, induced by the spin-orbit coupling are of the same order. Then diffusive modes corresponding to the total spin zero remain gapless only. Thus the number of triplet modes reduces to zero, $\mathcal{N}=0$.

For $\mathcal{N}=0$ the renormalization group flow of the interaction parameters $\gamma_{s,c}$ and the dimensionless resistance $t$ is governed by the following equations \cite{BurmistrovPRB2015}
\begin{subequations}
\begin{align}
    \frac{dt}{dy} & = - t^2(1+\gamma_s+2\gamma_c)/2 , \label{eq:N=0:RG:t}\\
    \frac{d}{dy}\begin{pmatrix} \gamma_s \\ \gamma_c \end{pmatrix} & = -\frac{t}{2} \begin{pmatrix} 1 & 2 \\ 1 & 0 \end{pmatrix} \begin{pmatrix} \gamma_s \\ \gamma_c \end{pmatrix}  .
    \label{eq:N=0:RG}
\end{align}
\end{subequations}
We emphasize that contrary to the case $\mathcal{N}=1$ there is the weak-antilocalization correction (term with unity in the brackets in the right hand side of Eq. \eqref{eq:N=0:RG:t}) that results in flow of $t$ towards zero. In what follows we neglect terms proportional to $\gamma_{s,c}$ (Altshuler-Aronov and density-of-states-type corrections) in Eq. \eqref{eq:N=0:RG:t} in comparison with weak antilocalization. 
In accordance with Eq. \eqref{eq:N=0:RG}, the interaction parameters flow approaching the BCS line $-\gamma_s = \gamma_c \equiv \gamma$. Projecting Eq. \eqref{eq:N=0:RG} onto the BCS line, we find
\begin{equation}
    d \gamma/dy = t\gamma/2, \;\;\; \gamma_0 = (2\gamma_{c0}-\gamma_{s0})/3 <0 .
    \label{eq:N=0:RG:BCS}
\end{equation}
Here $\gamma_0$ is the initial value of the effective attraction. We assume $t_0\gg|\gamma_0|$. Solving Eqs. \eqref{eq:N=0:RG:t} and \eqref{eq:N=0:RG:BCS}, we obtain
\begin{gather}
t(L) = \frac{t_0}{1+(t_0/2)\ln L/\ell}, \qquad \gamma(L) =\gamma_0 \frac{t_0}{t(L)}  .      \label{eq:N=0:Sol:RG}
\end{gather}
The effective attraction is growing with increase of the length scale. The superconducting critical temperature can be estimated from the condition $|\gamma(L_{T_c})|\sim t(L_{T_c})$ (see Appendix B). It results in the estimate \cite{BurmistrovPRL2012}
\begin{equation}
    \ln 1/(T_c\tau) \sim \frac{1}{\sqrt{|\gamma_0|t_0}} .
    \label{eq:N=0:Tc-RG}
\end{equation}
We note that the numerical factor cannot be determined reliably in this way.

\subsection{The critical temperature}

Let us now, in order to determine the superconducting transition temperature, solve the self-consistent equation \eqref{eq:N=1:sc-linearized}. It is convenient to introduce parametrization of the critical temperature in the form
\begin{equation}
T_c = (2\pi\tau)^{-1} \exp\left(4/t_0-4/t_c\right), 
\label{eq:Tc:par:N0}
\end{equation}
where, based on Eq. \eqref{eq:N=0:Tc-RG}, we expect $t_c$ to be of the order of $\sqrt{|\gamma_0| t_0}$. Hence Eq. \eqref{eq:N=1:sc-linearized} becomes
\begin{gather}
    \Delta_n = \frac{|\gamma_0|t_0}{4} \sum_{n'= 0}^{n_{\rm max}} \frac{4/t_c - \ln(n+n'+1)}{n'+1/2}  \Delta_{n'}. \label{eq:N=0:sc-Tc}
\end{gather}
Here $n_{\rm max}=1/(2\pi T_c\tau)\simeq \exp(4/t_c)$.
Again Eq. \eqref{eq:N=0:sc-Tc} 
can be considered as the maximal eigenvalue problem for the matrix $M_{nn'}(\zeta) = (\zeta - \ln(n+n'+1))/(n'+1/2)$ with $\zeta=4/t_c$. Numerical solution for the maximal eigenvalue reveals 
$\lambda_M\approx 0.41 (4/t_c)^2$
(see Appendix C). Thus we find
\begin{equation}
t_c\approx 1.3 \sqrt{|\gamma_0|t_0} .
\label{eq:N=0:Tc-num}
\end{equation}
The dependence of the right eigenvector corresponding to the maximal eigenvalue found numerically is shown in Fig. \ref{fig-Delta-near-Tc}. 


Similar to the case $\mathcal{N}=1$, we are able to solve the linearized self-consistency equation \eqref{eq:N=1:sc-linearized} analytically. 
Introducing the variable 
\begin{equation}
u_\varepsilon = \frac{4}{t(L_\varepsilon)} = \frac{4|\gamma(L_\varepsilon)|}{|\gamma_0| t_0}=\frac{4}{t_0} - \ln(\varepsilon \tau),
\end{equation}
replacing $\ln(n+n'+1)$ with $\ln(\max\{n+1/2,n'+1/2\})$, and using the Euler-Maclaurin formula, we get
\begin{gather}
    \frac{4 \Delta_{u_n}}{|\gamma_0| t_0} = u_n \int\limits_{u_{n}}^{u_{0}} du \Delta_u + \int\limits_{u_{\infty}}^{u_{n}} du u \Delta_u + a u_n \Delta_{u_0} . \label{eq:N=0:sc:Euler_Maclaurin}
\end{gather}
Here $a$ coincides with the one defined in Eq. \eqref{eq:N=1:a-def}. Next, writing $\Delta_{u_n} = \Delta_{u_0} f(u_n)$ with $f(u_0)=1$, we cast the above equation in the following differential equation,
\begin{gather}
    f^{\prime \prime}(u) = -(|\gamma_0| t_0/4) f(u), \notag \\
    f(u_0) = 1, \;\;f^\prime(u_\infty) = \frac{f(u_\infty)}{u_\infty}, \;\; f^\prime(u_0) = \frac{|\gamma_0| t_0}{4} .
\end{gather}
The latter can be elementary solved:
\begin{gather}
    f(u) = \frac{F_0(u)}{F_0(u_0)}, \;\;\;
    F_0(u) \simeq \sin\Bigl[u \sqrt{|\gamma_0| t_0/4}\Bigr ]  \label{eq:N=0:f(u)} 
\end{gather}
Again, in the above equation the solution to $f(u)$ is written in the lowest order in the small parameters $|\gamma_0| \ll t_0 \ll 1$. The last step is to find $u_0 = 4/t_c$ from the relation $f^\prime(u_0) = |\gamma_0| t_0/4$. Simple algebra yields
\begin{gather}
    u_0 \simeq \pi/\sqrt{|\gamma_0| t_0},
    \quad \Leftrightarrow \quad 
    t_c\simeq \frac{4}{\pi} \sqrt{|\gamma_0| t_0} .
\label{eq:N=0:Tc-EulerMaclaurin}
\end{gather}
We point out remarkable agreement between the numerical and the analytical results, Eq. \eqref{eq:N=0:Tc-num} and Eq. \eqref{eq:N=0:Tc-EulerMaclaurin}: 1.3 and $4/\pi$, respectively.

The numerical solution of Eq. \eqref{eq:N=0:sc-Tc} for $T_c$ with arbitrary values of $t_0$ is shown in Fig. \ref{figTc}. Analytical expression \eqref{eq:N=0:Tc-EulerMaclaurin} (marked with a black dotted line) gives correct asymptotic values of $\ln T_c/T_c^{(BCS)}$ in the regime of large ratio $t_0/|\gamma_0|$.

\subsection{The spectral gap function}

\subsubsection{The gap function near $T_c$}

At temperatures $T_c-T \ll T_c$ we use Eq. \eqref{eq:N=1:sc-nearTc} in order to find $\Delta_\varepsilon$. We parametrize the temperature $T$ by $t_T$, such that $T = (2\pi\tau)^{-1} \exp(4/t_0-4/t_T)$. After one writes $\Delta_\varepsilon=\Delta_0(T) r_n$, Eq. \eqref{eq:N=1:sc-nearTc} becomes
\begin{gather}
    \lambda_M r_n \simeq \sum_{n^\prime=0}^{n_{\max}} M_{nn^\prime} (4/t_T) \left[r_{n^\prime} - \frac{\frac{\Delta_0^2(T)}{8\pi^2 T^2_c} r_{n^\prime}^3}{(n^\prime+1/2)^2}  \right], \label{eq:N=0:sc-Tc-rn:0}
\end{gather}
Here, we remind, $\lambda_M$ denotes the maximal eigenvalue of the matrix $M_{nn^\prime}(4/t_c)$. Using the identity $M_{nn^\prime} (4/t_T)=M_{nn^\prime} (4/t_c)+(4/t_T-4/t_c)/(n^\prime+1/2)$ and approximation $4/t_T-4/t_c\approx(T_c-T)/T_c$, we rewrite Eq. \eqref{eq:N=0:sc-Tc-rn:0} as 
\begin{gather}
    \lambda_M(4/t_c) r_n \simeq \sum_{n^\prime=0}^{n_{\max}} M_{nn^\prime} (4/t_c) \left[r_{n^\prime} - \frac{\frac{\Delta_0^2(T)}{8\pi^2 T^2_c} r_{n^\prime}^3}{(n^\prime+1/2)^2}  \right]
    \notag \\
    + \frac{T_c-T}{T_c} \sum_{n^\prime=0}^{n_{\max}} \frac{r_{n^\prime}}{n^\prime+1/2} . \label{eq:N=0:sc-Tc-rn}
\end{gather}
When $T = T_c$ the gap function turns to zero. This  implies that $r_n$ are the components of the leading eigenvector of matrix $M_{nn^\prime}(4/t_c)$. Multiplying Eq. \eqref{eq:N=0:sc-Tc-rn} on the left eigenvector of  $M_{nn^\prime}(4/t_c)$, $l_n=r_n/(n+1/2)$, we retrieve the result \eqref{eq:N=1:Tc-Delta0(T)} with $b_{\mathcal{N}}$ for $\mathcal{N} = 0$ being
\begin{gather}
        b_0  = \frac{7\zeta(3)}{\lambda_M} \frac{\left(\sum\limits_{n}^{n_{\max}} r_{n}/(n + 1/2) \right)^2}{\sum\limits_{n=0}^{n_{\max}} r_{n}^4/(n+1/2)^3 } .
        \label{eq:N=0:Tc-b}
\end{gather}
As in the case of $\mathcal{N}=1$ one can check that within our approximation $b_0=1$ as in the BCS theory. The denominator in the right hand side of Eq. \eqref{eq:N=0:Tc-b} is treated in the same fashion as in the case of $\mathcal{N} = 1$. While in the numerator we write $r_n = f(u_n)$, where $f(u)$ is given in Eq. \eqref{eq:N=0:f(u)}. Using $F_0(u) = \sin(u/\sqrt{\lambda_M})$ and $u_0 = \sqrt{\lambda_M} \pi/2$, one immediately finds
\begin{gather}
    \left(\sum\limits_{n=0}^{n_{\max}} \frac{r_n}{n+1/2} \right)^2 \simeq \left( \int\limits_{0}^{u_0} du \; f(u) \right)^2 = \lambda_M.
\end{gather}
Therefore, combining the results for numerator and denominator we restore $b_0=1$.

\subsubsection{The gap function at 
$T \ll T_c$}

The spectral gap function at $T=0$ can be found in the same way as it was done in Sec. \ref{Sec:gapT0:N1} for $\mathcal{N}=1$. After straightforward calculations, we find 
\begin{equation}
    \Delta_\varepsilon \simeq \Delta_0 
    \begin{cases}
        1, & \varepsilon \lesssim \Delta_0, \\
        F_0(u_\varepsilon)/F_0(u_{\Delta_0}), & \varepsilon \gtrsim \Delta_0 .
    \end{cases} 
\end{equation}
The maximal magnitude of the spectral gap, $\Delta_0$, is given by  
the expression similar to Eq.~\eqref{eq:Tc:par:N0},
\begin{equation}
\Delta_0 = (2\pi\tau)^{-1} e^{4/t_0-4/t_{\Delta_0}}, \quad 
t_{\Delta_0} \sim \sqrt{|\gamma_0|t_0} .
\end{equation}
Unfortunately, within our approximation we cannot unambiguously determine the numerical factor in the ratio $t_{\Delta_0}/ \sqrt{|\gamma_0|t_0}$ since it requires knowledge of   
the precise dependence of the infrared lengthscale $L_{E_\varepsilon+E_{\varepsilon^\prime}}$ on $\Delta_\varepsilon$ and $\Delta_{\varepsilon^\prime}$, see Eq. \eqref{eq:NSLM:sc-general}. Since $t_{\Delta_0}$ stands in the exponent of the expression for $\Delta_0$, we cannot exclude a possibility that $\Delta_0$ differs parametrically from $T_c$.  

We note that the superconducting order parameter (which coincides with the spectral gap function at energies $\varepsilon\sim 1/\tau$) is proportional to $\Delta_0\sqrt{|\gamma_0|/t_0}\ll\Delta_0$.
Typical dependence of $\Delta_\varepsilon$ on $\varepsilon$ is shown in Fig.~\ref{fig1}.

The change of $\Delta_\varepsilon$ with increasing temperature is the same as in the case of $\mathcal{N}=1$. At $T\ll T_c$ the amplitude $\Delta_0$ is decreasing in accordance with Eq. \eqref{eq:Delta:T}. 

\section{Local density of states\label{Sec:LDOS}}

In this section we discuss  the local density of states and its mesoscopic fluctuations in the superconducting state. For the sake of simplicity we consider $T=0$.

The disorder-averaged density of states can be found from the solution of the Usadel equation
\begin{align}
    \langle\rho(E)\rangle & = 
\nu \Re \cos\theta_{\varepsilon \to -iE + 0}   \notag \\&  =
    \nu \Re \frac{\varepsilon}{\sqrt{\varepsilon^2+\Delta_\varepsilon^2}} \Big|_{\varepsilon \to -iE + 0}.
\end{align}
\noindent Here the analytical continuation from Matsubara energies to real energies is performed, $i\varepsilon \to E + i0$.

The behavior of the density of states at energies $E$ close to $\Delta_0$ depends on fine structure of $\Delta_\varepsilon$ at $\varepsilon\sim \Delta_0$. Thus we have no access to these range of energies. The only statement is possible to make is the existence of the spectral gap of the order of $\Delta_0$. Away from $\Delta_0$, i.e. at $E\gtrsim \Delta_0$, using the smallness of $\Delta_\varepsilon/\varepsilon$, we find 
\begin{gather}
    \frac{\langle\rho(E)\rangle}{\nu} \simeq 1 + \Re\frac{\Delta^2(E)}{2E^2} 
    \simeq 1+ \frac{\Delta_0^2}{2E^2}\frac{F^2_\mathcal{N}(u_E)}{F^2_\mathcal{N}(u_{\Delta_0})} .
\end{gather}
Here $u_E=u_{\Delta_0}(E/\Delta_0)^{-t_0/2}$ and $u_E=u_{\Delta_0}-\ln(E/\Delta_0)$ for $\mathcal{N}=1$ and $0$, respectivelty.

Now we estimate mesoscopic fluctuations of the local density of states. This can be done using the nonlinear sigma model approach in a way similar to the one of Ref. \cite{BurmistrovAnnPhys2021}. 
We restrict our consideration by energy range
\begin{equation}
  \Delta_0 \ll \frac{1}{\tau}e^{-2(2-\mathcal{N})/t_0}  \ll E \lesssim  \frac{1}{\tau} . \label{eq:energy_range}
\end{equation}
In this range we can neglect the energy dependence of $\Delta_\varepsilon$ and approximate it by its non-renormalized value $\Delta$. Then we obtain for the variance of the local density of states the following result (see Appendix D)
\begin{gather}
    \frac{\langle [\delta \rho(E, \bm{r})]^2 \rangle}{\nu^2} = \frac{ 1+\mathcal{N}}{g} \re \int \frac{d^2 \bm{q}}{(2\pi)^2} 
    \bigg [ \frac{2E^2-\Delta^2}{E^2-\Delta^2}\frac{1}{q^2} 
    \notag \\
    +\frac{\Delta^2}{E^2-\Delta^2} \frac{D}{Dq^2 + 2 i \sqrt{E^2- \Delta^2}} \bigg] .
    \label{eq:varLDoS:1}
\end{gather}
We note that the contribution in the first line of Eq. \eqref{eq:varLDoS:1} corresponds to correlations between electron-like and electron-like excitations. They do not feel the superconducting spectral gap. This explains why the infrared divergence in diffusive propagator persist inside superconductor. The contribution in the second line of Eq. \eqref{eq:varLDoS:1} is due to correlations between electron-like and hole-like excitations which are split by the 
twice the superconducting gap.

Performing integration over momentum and expansion in $\Delta/E\ll 1$, we find
\begin{gather}
    \frac{\langle [\delta \rho(E, \bm{r})]^2 \rangle}{\langle \rho(E)\rangle^2} = \frac{(1+\mathcal{N})t_0}{2} \Bigl (
    \ln \frac{L}{\ell}+\frac{\Delta^2}{2E^2}
    \ln \frac{L_E}{\ell}\Bigr ) .
    \label{eq:varLDOS:2}
\end{gather}
We emphasize logarithmic divergence of the variance with the system size $L$. This signals about strong mesoscopic fluctuations of the local density of states in disordered superconductors with the spin-orbit coupling similar to the case when the spin-orbit coupling is absent \cite{BurmistrovAnnPhys2021,Stosiek2021}.

The renormalization of diffusive propagator ignored so far results in substitution of $L$ by $\min\{L,L_E^{(\phi)}\}$ where $L_E^{(\phi)}$ denotes the dephasing length induced by electron-electron interactions. Unfortunately, at present, there is no complete theory of dephasing rate in disordered superconductors. Using the results of Ref. \cite{Belitz1991}, one can estimate dephasing length due to electron-electron interaction at $E\gg\Delta$ as $L_\Delta/(|\gamma_0| \sqrt{t_0}) \gg L_\Delta$. We note that such estimate is applicable at $T\ll T_c$. Close to superconducting transition the dephasing rate is enhanced due to the superconducting fluctuations (for details see Ref.~\cite{Woelfle1985}).

With the help of the renormalization group we can convert the perturbative, infrared divergent, result \eqref{eq:varLDOS:2} into the result for the second moment of the local density of state:
\begin{equation}
    \frac{\langle \rho^2(E, \bm{r}) \rangle}{\langle \rho(E)\rangle^2}
    =\begin{cases}
    (\min\{L,L_E^{(\phi)}\}/\ell)^{t_0}, & \mathcal{N}=1 ,\\
    1+ (t_0/2) \ln (\min\{L,L_E^{(\phi)}\}/\ell), 
    &  \mathcal{N}=0 .
    \end{cases}
    \label{eq:varLDOS:3}
\end{equation}

One can also generalize expression \eqref{eq:varLDOS:2} to the pair correlation function of the local density of states at differing energies $E\gg \Delta$ and $E^\prime = E + \omega \gg \Delta$ (see Appendix D), 
\begin{gather}   
\frac{\langle \delta \rho(E, \bm{r}) \delta \rho(E^\prime, \bm{r}) \rangle}{\langle \rho(E) \rangle \langle \rho(E^\prime) \rangle} \simeq \frac{(1+\mathcal{N})t_0}{2} \ln \frac{\min\{L,L_E^{(\phi)},L_\omega\}}{\ell} . \label{eq:MFLDoS:K2:E(E+w)}
\end{gather}
For the autocorrelation function of moments one can use Eq. \eqref{eq:varLDOS:3} in which $\min\{L,L_E^{(\phi)}\}$ should be substituted by $\min\{L,L_E^{(\phi)},L_\omega\}$.

We note that one can compute the disorder-averaged higher moments of the local density of states. Similarly to the case without spin-orbit interaction \cite{BurmistrovAnnPhys2021}, the moments correspond to the log-normal distribution for the local density of states.

\section{Summary and conclusions\label{Sec:DiscConc}}

To summarize, we have developed the theory of the multifractally-enhanced superconducting state in thin films with spin-orbit coupling
under assumption of the presence of weak short-ranged electron-electron interaction.
We considered the case of Ising spin-orbit coupling ($\mathcal{N}=1$) for which renormalization of the normal state resistance is negligible and the case of strong spin-orbit coupling ($\mathcal{N}=0$) in which renormalization of $t$ is dominated by weak-antilocalization correction. Together with the case without spin-orbit coupling ($\mathcal{N}=3$) studied in Ref. \cite{BurmistrovAnnPhys2021}, we have now the theory for all three possible behavior of the normal state resistance with the system size: increasing, decreasing and constant.  

Following approach of Ref. \cite{BurmistrovAnnPhys2021}, we treated the fluctuations around the mean-field spatially homogeneous saddle-point and derived the modified Usadel equation and self-consistency equation that incorporates the interplay of disorder and interactions at high energies.
The derived equations allows us to determine accurately the superconducting transition temperature. The later is enhanced (even for ($\mathcal{N}=0$) in comparison with the BCS result in the absence of disorder. The solution of the modified Usadel equation and self-consistency equation yields $T_c$ which coincides parametrically with the estimate from the Cooper-channel instability in renormalization group equations for the normal phase. It is worth comparing the multifractally increased critical temperatures for different number of triplet modes. In the absence of spin-orbit coupling, see Ref. \cite{BurmistrovAnnPhys2021}, superconducting phase is expected to be present for temperatures lower than $T_c^{\mathcal{N} = 3} \sim \tau^{-1} \exp(-2/t_0)$. In contrast,  we predict $T_c^{\mathcal{N} = 1} \sim \tau^{-1} \exp(-(2/t_0) \ln (t_0/(1.4 |\gamma_0|)))$ for $\mathcal{N} = 1$ and $T_c^{\mathcal{N} = 0} \sim \tau^{-1} \exp(-3.1/\sqrt{|\gamma_0| t_0})$ for $\mathcal{N} = 0$, respectively. Therefore, $T_c$ increases with the number of triplet diffusive modes: $T_c^{\mathcal{N} = 3} \gg T_c^{\mathcal{N}= 1} \gg T_c^{\mathcal{N} = 0}$. 

The disorder-induced energy dependence of the effective attraction translates into the energy dependence of the spectral gap function. The form of this energy dependence is sensitive to the number of triplet modes $\mathcal{N}$. In the case $\mathcal{N}=1$ the energy dependence of the spectral gap is concave (the same holds for $\mathcal{N}=3$), see Fig. \ref{fig1}. For $\mathcal{N}=0$ $\Delta_\varepsilon$ becomes convex function of $\varepsilon$. The interplay of disorder and interactions lead to parametrical enhancement of the spectral gap function at low energies in comparison with its magnitude at energies $\sim 1/\tau$. 

In spite of such strong distinction from the BCS theory in which the spectral gap function is independent of the energy, the basic relations for the BCS model: behavior of the gap just below $T_c$ and at low temperatures $T\ll T_c$, remain untouched. This indicates that one can use the BCS theory in order to describe the temperature dependence of the spectral gap in weakly disordered films.      
In the presence of spin-orbit coupling we revealed strong mesoscopic fluctuations of the local density of states in the superconducting state. These fluctuations persist upto the lengthscale set by the dephasing length. The presence of spin-orbit coupling reduces the amplitude of the variance of the local density of states by a factor 2 (for $\mathcal{N}=1$) and by a factor 4 (for $\mathcal{N}=0$). Due to different energy dependence of the gap function for different $\mathcal{N}$, the energy dependence of the variance is sensitive to the number of massless triplet modes. The most pronounced difference is expected near the coherence peak, $E\sim \pm \Delta_0$. However this region is beyond the accuracy of our calculations.

It is instructive to compare the fluctuations of the local density of states with the fluctuations of the superconducting order parameter. We note that although our approach assumes spatially constant order parameter one can study its mesoscopic fluctuations. In the case of the presence of the spin-orbit coupling, we find 
\begin{equation}
\frac{\langle (\delta\Delta)^2\rangle}{\Delta^2} \simeq\frac{(1+\mathcal{N})t_0}{2}\ln \frac{L_{\Delta_0}}{\ell} .
\label{eq:SOP}
\end{equation}
We emphasize the close similarity between Eq. \eqref{eq:MFLDoS:K2:E(E+w)} and \eqref{eq:SOP}. However, there is a crucial difference: the mesoscopic fluctuations of the superconducting order parameter are controlled by the coherence length $L_{\Delta_0}$ in the infrared. Since $L_E^{(\phi)} \gg L_{\Delta_0}$ we expect $\langle (\delta \rho)^2\rangle/\langle\rho\rangle^2 \gg \langle (\delta\Delta)^2\rangle/\Delta^2$. 
For $\mathcal{N}=1$ ($\mathcal{N}=0)$ we can estimate $\langle (\delta\Delta)^2\rangle/\Delta^2$ as $\sim\ln(t_0/|\gamma_0|)$ ($\sim\sqrt{t_0/|\gamma_0|}$), respectively.

The results presented in this paper can be generalized in several directions. Most interesting to study renormalization of the diffusive propagators and to formulate the equation for the spectral gap in a fully self-consistent way. With such equation in hands one can determine $\Delta_0$ upto the numerical factor in order to compare it with the expression for $T_c$. Also such computation for the diffusive propagators  allows to determine the dephasing length serving a cut-off for the mesoscopic fluctuations of the local density of states.   
Further our theory can be extended to include the Coulomb interaction. Additionally, it would be interesting to go beyond weak-coupling for superconductivity and to study  the multifractal effects at the BCS – BEC crossover \cite{Loh2016,Kagan2021}. 
Also our work can be extended to consider systems with singular dynamical interaction between electrons, see Refs. \cite{GammaModI,GammaModII,GammaModIII,GammaModIV,GammaModV,GammaModVI}, in the presence of disorder. This can be achieved along the approaches of Refs. \cite{Nosov2020,Foster2022}.

Finally, we mention that our theory ignores phase fluctuations of the order parameter. The latter are known to be responsible for the Berezinskii- Kosterlitz-Thouless transition in superconducting films.
Such fluctuations can be taken into account in the way similar to the one in Refs. \cite{Konig2015,Konig2021}. However, for weakly disordered superconducting films effects related with phase fluctuations are expected to be weak \cite{Beasley1979,Konig2015}.

Authors are grateful to C. Brun, T. Cren, F. Evers, I. Gornyi, M. Lizee, A. Mirlin, P. Nosov, S. Raghu, and M. Stosiek for collaboration on related projects and useful discussions. The research is partially supported by the Russian Foundation for Basic Research (Grant No. 20-52- 12013) 
and by the Basic Research Program of HSE.


\appendix
\section{Finkel'stein nonlinear sigma model}
\label{App-NLSM}

In this Appendix we present details of the Finkel'stein nonlinear sigma model formalism. 

The action of an electron liquid in a disordered metal with spin-orbit coupling is given by
\begin{gather}
    S = S_\sigma + S_{\text{int}}^{(\rho)}+S_{\text{int}}^{(\sigma)}+S_{\text{int}}^{(c)}+S_{\text{so}}, \label{eq:AppA:action}
\end{gather}

\noindent where the first term comes from non-interacting fermions. The next three terms correspond to electron-electron interactions in the particle-hole singlet channel, $S_{\text{int}}^{(\rho)}$, in the particle-hole triplet channel, $S_{\text{int}}^{(\sigma)}$, and  in the particle-particle channel $S_{\text{int}}^{(c)}$. The last term appears due to spin-orbit coupling. The above mentioned contributions reads (see Refs. \cite{Fin,KB,Burmistrov2019} for review)
\begin{subequations}
\begin{align}
    S_\sigma = & - \frac{g}{32}\int_{\bm{r}} \Tr(\nabla Q)^2 + 2 Z_\omega \int_{\bm{r}}\Tr \hat{\varepsilon} Q \\
    S_{\text{int}}^{(\rho)} = & -\frac{\pi T}{4} \Gamma_s \sum_{r=0,3} \sum_{\alpha,n} \int_{\bm{r}} \Tr I^\alpha_n t_{r0} Q \Tr I^\alpha_{-n} t_{r0} Q \\ 
    S_{\text{int}}^{(\sigma)} = & -\frac{\pi T}{4} \Gamma_t \sum_{\substack{r=0,3 \\ j = 1,2,3}}\sum_{\alpha,n} \int_{\bm{r}} \Tr I^\alpha_n t_{rj} Q \Tr I^\alpha_{-n} t_{rj} Q\\
    S_{\text{int}}^{(c)} = & -\frac{\pi T}{4}\Gamma_c \sum_{r=1,2} \sum_{\alpha,n} \int_{\bm{r}} \Tr t_{r0} L^\alpha_n Q \Tr t_{r0} L^\alpha_n Q  \label{eq:AppA:S_int^c} \\
    S_{\text{so}} = & \frac{\pi \nu}{2} \sum_{j=1,2,3} \frac{1}{\tau_{\rm so}^{(j)}} \int_{\bm{r}} \Tr(t_{0j} Q)^2
\end{align}
\end{subequations}
In what is written above $g$ is the total Drude conductivity (in units $e^2/h$ and including spin). The parameter $Z_\omega$ describes the renormalization of the frequency term \cite{Fin}. Its bare value is given as $\pi \nu/4$. Interaction amplitudes in the singlet, the triplet, and the particle-hole channels are designated as $\Gamma_s$, $\Gamma_t$, and $\Gamma_c$, respectively. It is convenient to introduce the dimensionless interaction parameters $\gamma_{s,t,c} \equiv \Gamma_{s,t,c}/Z_\omega$. $1/\tau_{\text{so}}^{(j)}$ stands for the spin-orbit scattering rates.

The matrix field $Q(\bm{r})$ and the trace $\Tr$ operate in the replica ($\alpha, \beta$), Matsubara ($n,m$), spin ($j=0,1,2,3$), and particle-hole ($r=0,1,2,3$) spaces. The matrix field $Q(\bm{r})$ obeys the nonlinear constraint, as well as the charge-conjugation symmetry relation
\begin{equation}
    Q^2 = 1, \;\;\; \Tr Q = 0,\;\;\; Q = Q^\dag = -C Q^T C, \label{eq:AppA:constraints}
\end{equation}

\noindent where $C = i t_{12}$ and the matrix $t_{rj}$ is
\begin{equation}
    t_{rj} = \tau_r \otimes s_j, \quad r,j=0,1,2,3 .
\end{equation}

\noindent In the expression above $r$ and $j$ subscripts correspond to particle-hole and spin spaces, respectively. $\tau_r$ and $s_j$ denote standard Pauli matrices,
\begin{align}
    \tau_0 / s_0 = \begin{pmatrix} 1 & 0 \\ 0 & 1 \end{pmatrix}, & \quad \tau_1 / s_1 = \begin{pmatrix} 0 & 1 \\ 1 & 0 \end{pmatrix}, \notag \\
    \tau_2 / s_2 = \begin{pmatrix} 0 & -i \\ i & 0 \end{pmatrix}, & \quad \tau_3 / 
    s_3 = \begin{pmatrix} 1 & 0 \\ 0 & -1 \end{pmatrix}.
\end{align}

\noindent Taking constraints \eqref{eq:AppA:constraints} into account, we use the following parametrization for the matrix field $Q(\bm{r})$
\begin{gather}
    Q = U^{-1} \Lambda U, \;\;\; U^\dag = U^{-1}, \;\;\; CU^T=U^{-1}C, \notag \\
    \Lambda_{nm}^{\alpha \beta} = \sgn \varepsilon_n \delta_{\varepsilon_n, \varepsilon_m} \delta^{\alpha \beta} t_{00}
\end{gather}

\noindent The constant matrices in the action \eqref{eq:AppA:action} are given by the following expressions ($\omega_k = 2\pi T k $): 
\begin{gather}
    \hat{\varepsilon}^{\alpha \beta}_{nm} = \varepsilon_n \delta_{\varepsilon_n, \varepsilon_m} \delta^{\alpha \beta} t_{00}, \notag \\
    (I^{\gamma}_k)^{\alpha \beta}_{nm} = \delta_{\varepsilon_n - \varepsilon_m, \omega_k}\delta^{\alpha \beta}\delta^{\alpha \gamma} t_{00}, \\
    (L^{\gamma}_k)^{\alpha \beta}_{nm} = \delta_{\varepsilon_n +\varepsilon_m, \omega_k}\delta^{\alpha \beta}\delta^{\alpha \gamma} t_{00}.\notag
\end{gather}

Following Ref. \cite{BurmistrovAnnPhys2021}, we employ the above technique to describe the low-energy physics in the broken symmetry superconducting state. In order to do so one needs to single out the static term with $n = 0$ in the particle-particle channel, see Eq. \eqref{eq:AppA:S_int^c}, and introduce two decoupling fields $\Delta_r^\alpha (\bm{r})$ with $r = 1,2$. Upon Hubbard–Stratonovich transformation one finds
\begin{gather}
    S_{\text{int}}^{(c)} =\sum_{r=1,2}  \int_{\bm{r}} \Biggl[  \frac{4 Z_\omega}{\pi T \gamma_c}  [\Delta_r^\alpha(\bm{r})]^2 
    + 2 Z_\omega  \Delta_r^\alpha (\bm{r}) \Tr t_{r0} L_0^\alpha Q \notag \\ 
    -\frac{\pi T}{4}\Gamma_c \sum_{n \neq 0}  \left(\Tr t_{r0} L^\alpha_n Q\right)^2\Biggr ] .
\end{gather}


Variation of the total action with respect to $Q(\bm{r})$ and $\Delta_r^\alpha(\bm{r})$ give rise to the Usadel equation and the self-consistency relations for $\Delta_r^\alpha (\bm{r})$, $r = 1,2$. In turn, these equations may generate many spatially dependent solutions. In order to account for them we assume $1/g \ll 1$ and exploit the renormalization group technique by treating the spatially dependent solutions $Q(\bm{r})$ as fluctuations around some spatially independent solution $\underline{Q}$.

This program can be performed as follows. At first, we distinguish spatially independent and spatially dependent components of the fields $\Delta_r^\alpha(\bm{r})$, $r = 1,2$, they read
\begin{equation}
    \Delta_r^\alpha(\bm{r})  = \underline{\Delta}_r^\alpha+\delta \Delta_r^\alpha(\bm{r}),
    \quad  \int_{\bm{r}} \ \delta \Delta_r^\alpha(\bm{r}) = 0 .
\end{equation}

We point out that fluctuations of the order parameter are now contained in $\delta \Delta_r^\alpha(\bm{r})$. On the other hand, one can perform a formally exact integration over the fields $\delta \Delta_r^\alpha(\bm{r})$ in the action. The latter transfers information about fluctuations of the order parameter entirely onto the field $Q(\bm{r})$. Accordingly, we get
\begin{gather}
    S_{\rm int}^{(c)} =  2 Z_\omega  V \sum_{\alpha}\sum_{r=1,2}  \Biggl \{ 
    \underline{\Delta}_r^\alpha \Tr t_{r0} L_0^\alpha \overline{Q}
    +\frac{2}{\pi T \gamma_c}   
    \bigl [ \underline{\Delta}_r^\alpha \bigr ]^2 \Biggr \}\notag \\
    + \hat S_{\rm int}^{(c)} ,
\end{gather}
where $V$ is the volume of a superconductor, 
\begin{equation}
    \overline{Q} = \frac{1}{V} \int_{\bm{r}}\, Q(\bm{r}) ,
\end{equation}
and
\begin{equation}
    \hat S_{\rm int}^{(c)} 
    =  -\frac{\pi T}{4}  \Gamma_c \sum_{\alpha,n} \sum_{r=1,2}  \int_{\bm{r}}\bigl( \Tr  t_{r0} L_n^\alpha Q_n \bigr )^2\ .
    \label{eq:hatSc}
\end{equation}
Where $Q_n=Q-\overline{Q} \delta_{n,0}$.


The described above procedure leads to a new saddle-point equation for $Q(\bm{r})$ and self-consistency equations for $\underline{\Delta}_r^\alpha $,
\begin{gather}
D\nabla (Q\nabla Q) -[\hat\varepsilon+\hat{\underline{\Delta}},Q] 
 + \frac{\pi T}{4}  \sum_{\alpha,n}\Biggl [ \sum_{r=1,2} \gamma_c  [t_{r0} L_n^\alpha, Q] 
\notag \\
\times \Tr  t_{r0} L_n^\alpha Q_n 
 + \sum_{r=0,3} \sum_{j=0}^3 \gamma_j
[I_{-n}^\alpha t_{rj}, Q] \Tr I_n^\alpha t_{rj} Q \Biggr ]
 =0 ,
 \notag
\\
     \underline{\Delta}_r^\alpha = \frac{\pi T}{4} |\gamma_c| \Tr t_{r0} L_0^\alpha \overline{Q}, \quad r =1,2 \ .  
     \label{eq:SCE:full:1} 
\end{gather}

Now, one can investigate solutions of the latter equations. In the mean-field description one ignores fluctuations, seeking a spatially independent solution solely. This solution can be conveniently parametrized with the so-called spectral angle, $\theta_{\varepsilon_n}$, which is function of Matsubara energies $\varepsilon_n$. In terms of the spectral angle the saddle-point solution reads 
\begin{gather}
 \underline{Q} = R^{-1}\Lambda R, \qquad 
    R_{nm}^{\alpha\beta} =
    \Bigl [ \delta_{\varepsilon_n,\varepsilon_m}
    \cos \frac{\theta_{\varepsilon_n}}{2} , \notag \\ -  t_{\phi} \delta_{\varepsilon_n,-\varepsilon_m} \sgn \varepsilon_m \sin \frac{\theta_{\varepsilon_n}}{2} \Bigr] \delta^{\alpha\beta}
\notag \\
     t_\phi =\cos\phi \ t_{10}+\sin\phi\ t_{20} , \notag \\
     \underline{\Delta}_1^\alpha=\Delta \cos\phi, \qquad \underline{\Delta}_2^\alpha =\Delta \sin \phi .
 \label{eq:sol:sp}
\end{gather}

Substituting the expressions for $\underline{Q}$ and $\underline{\Delta}_r^\alpha$ from \eqref{eq:sol:sp} into the Usadel and self-consistent equations, we find
\begin{subequations}
\begin{gather}
    \frac{D}{2} \nabla^2 \theta_{\varepsilon_n} - |\varepsilon_n|\sin \theta_{\varepsilon_n}+\Delta \cos\theta_{\varepsilon_n}
    = 0 ,
\label{eq:Usadel:theta} \\
  \Delta = \pi  T |\gamma_c|\sum_{\varepsilon_n} \sin \theta_{\varepsilon_n} .
\label{eq:SCE:theta} 
\end{gather}
\end{subequations}
The spatially independent solution of Eq. \eqref{eq:Usadel:theta} reduces 
Eq. \eqref{eq:SCE:theta} to the usual self-consistent equation of the BCS theory. Thus $T_c$ becomes insensitive to disorder in accordance with the "Anderson theorem". 


However, we are interested in more intricate picture, when fluctuations of $Q(\bm{r})$ around the saddle-point solution $\underline{Q}$ are taken into consideration. The latter corresponds to the interaction of the diffusive modes. We use the square-root parametrization of the matrix field to get the perturbative expansion of $Q(\bm{r})$ field around the saddle-point solution, 
\begin{gather}
    Q = R^{-1} \Bigl( W +\Lambda \sqrt{1-W^2}\Bigr ) R,  \quad 
    W= \begin{pmatrix}
    0 & w\\
    \overline{w} & 0
    \end{pmatrix} . \label{eq:Q-W}
\end{gather}
Here $W$-field satisfies the charge-conjugation constraints:
\begin{gather}
    \overline{w} = -C w^T C,\qquad w = - C w^* C .
\end{gather}

Before moving on to fluctuations, we also need to know the propagators for diffusive modes. 
We limit ourselves to the lowest order in residual electron-electron interactions, which corresponds to small magnitudes of the bare interaction parameters $|\gamma_{s0,t0,c0}| \ll 1$. Within this approximation, one finds
\begin{gather}
    \Bigl \langle [w_{rj}(\bm{p})]^{\alpha_1\beta_1}_{n_1n_2} [\overline{w}_{rj}(-\bm{p})]^{\beta_2\alpha_2}_{n_4n_3} 
    \Bigr \rangle 
    = \frac{2}{g} \delta^{\alpha_1\alpha_2}
     \delta^{\beta_1\beta_2}
    \delta_{\varepsilon_{n_1},\varepsilon_{n_3}} 
    \notag \\
    \times
    \delta_{\varepsilon_{n_2},\varepsilon_{n_4}}
    \mathcal{D}_p(i\varepsilon_{n_1},i\varepsilon_{n_2}) , \notag \\
    \mathcal{D}_p(i\varepsilon_{n_1},i\varepsilon_{n_2}) 
    =\frac{1}{p^2+ E_{\varepsilon_{n_1}}/D+E_{\varepsilon_{n_2}}/D}  ,
\label{eq:prop:Dp}
\end{gather}
where we remind $E_{\varepsilon_{n}} = |\varepsilon_{n}| \cos \theta_{\varepsilon_{n}} +
    \Delta\sin \theta_{\varepsilon_{n}}$. 
It is crucial to keep in mind here that the result \eqref{eq:prop:Dp} ignores spin-orbit term in the action, $S_{\rm so}$. It leads to appearance of additional mass term (proportional to $1/\tau_{\text{so}}$) in the denominator of the diffusive propagators \eqref{eq:prop:Dp}. 
Thus the corresponding modes will be suppressed in the diffusive limit. In other words, in the above equations $j$ accounts only for the gapless modes, i.e. $j = 0$ for $\mathcal{N}=0$ and $j=0,3$ for $\mathcal{N}=1$.

Hereby we are all set to investigate how the fluctuations of $Q$ renormalizes the NLSM action. To the lowest order in disorder we approximate $Q$ as 
$Q \simeq \underline{Q} + R^{-1} W R$ .
This yields the following fluctuation correction from the interacting part of the action (in the particle-hole channel),
\begin{gather}
    S_{\rm int}^{(\rho)}{+}S_{\rm int}^{(\sigma)} 
    {\to}
    -\frac{\pi T}{4}\!\int\limits_{\bm{r}} \!  \sum_{\alpha,n}\sum_{r=0,3}\sum_{j} \Gamma_j  \langle  \Tr \bigl [ R I_n^\alpha t_{rj} R^{-1} W\bigr ] \notag \\
    \times
    \Tr \bigl [ R I_{-n}^\alpha t_{rj} R^{-1} W \bigr ]  \rangle .
\end{gather}

Here and later a summation is performed over $j = 0$ for $\mathcal{N}=0$ and over $j = 0,3$ for $\mathcal{N}=1$. Given the expression for the propagators \eqref{eq:prop:Dp}, we obtain (for more details see Ref. \cite{BurmistrovAnnPhys2021})
\begin{gather}
    {}\hspace{-2cm} S_{\rm int}^{(\rho)}+S_{\rm int}^{(\sigma)} 
    \to
     \frac{32\pi T N_r V}{g} (\Gamma_s-\mathcal{N}\Gamma_t) \notag \\    
     \times \sum_{\varepsilon,\varepsilon^\prime>0} \sin \theta_{\varepsilon} \sin \theta_{\varepsilon^\prime} 
     \int_q \mathcal{D}_q(i\varepsilon,-i\varepsilon^\prime)\ ,
\label{eq:ren:Srho+sigma}
\end{gather}
where  $\int_q\equiv\int d^2 \bm{q}/(2\pi)^2$. In its turn, the interaction in the Cooper channel renormalizes as follows
\begin{gather}
    \hat{S}_{\rm int}^{(c)} 
    \to -\frac{32 \pi T \Gamma_c N_r}{g} \sum_{\varepsilon>0}  \mathcal{D}_{q=0}(i\varepsilon,-i\varepsilon) \sin^2 \theta_\varepsilon .
\label{eq:ren:Scooper1}
\end{gather}

Together Eqs. \eqref{eq:ren:Srho+sigma} and \eqref{eq:ren:Scooper1} contribute to the following modified action
\begin{gather}
    S[\underline{Q}] \to 16\pi T Z_\omega N_r V \Biggl \{
    \frac{\Delta^2}{4\pi T \gamma_c} 
        + \sum_{\varepsilon>0} \Bigl [ \varepsilon \cos\theta_{\varepsilon} + \Delta \sin \theta_{\varepsilon}\Bigr ] 
        \notag \\
        +\frac{2\pi T(\gamma_s-\mathcal{N}\gamma_t)}{g}   \sum_{\varepsilon,\varepsilon^\prime>0} \sin \theta_{\varepsilon} \sin \theta_{\varepsilon^\prime} 
     \int_q \mathcal{D}_q(i\varepsilon,-i\varepsilon^\prime)
    \Biggl \}\ .
\label{eq:eff:pot}
\end{gather}
Finally, variation of Eq. \eqref{eq:eff:pot} with respect to $\theta_\varepsilon$ and $\Delta$, leads to Eqs. \eqref{eq:NLSM:Usadel} and \eqref{eq:Delta:eps}, accordingly.

\section{Critical temperature}
\label{App-Tc-RG}

In this appendix we estimate the critical temperature using the renormalization group equations describing the renormalization of resistivity and interactions in the normal phase. The full set of one-loop (lowest order in disorder) renormalization group equations has been derived by means of the background field renormalization of the Finkelstein nonlinear sigma model, see Eqs. (47)-(51) in Ref. \cite{BurmistrovPRB2015}. Expanding these equations in $|\gamma_{s,t,c}|\ll 1$ and choosing $\mathcal{N} = 1$, we find 
\begin{subequations}
\begin{gather}
    \frac{dt}{dy} = - t^2 (\gamma_s +\gamma_t +2 \gamma_c])/2,  \label{eq:N=1:t-RG:a} \\
    \frac{d}{dy}\begin{pmatrix} \gamma_s \\ \gamma_t \\ \gamma_c \end{pmatrix} = - \frac{t}{2}
    \begin{pmatrix}
    1 & 1 & 2 \\
    1 & 1 & -2 \\
    1 & -1 & 0
    \end{pmatrix}
    \begin{pmatrix} \gamma_s \\ \gamma_t \\ \gamma_c \end{pmatrix} - \begin{pmatrix} 0 \\ 0 \\ 2\gamma_c^2 \end{pmatrix} . \label{eq:N=1:RG:a}
\end{gather}
\end{subequations}

\noindent Equation \eqref{eq:N=1:t-RG:a} implies that dimensionless resistance $t$ remains constant 
and equals to its bare value $t_0$. Projecting Eqs. \eqref{eq:N=1:RG:a} onto the BCS line $-\gamma_s = \gamma_t = \gamma_c = \gamma$ we obtain 
\begin{gather}
    \frac{d\gamma}{dy} = t_0 \gamma - \gamma^2, \quad \gamma_0 = (\gamma_{t0}-\gamma_{s0}+2\gamma_{c0})/4 <0  . 
\end{gather}
After solving the equation written above, we find that renormalization group flow diverges at $y_c = t_0^{-1} \ln(1+ t_0/|\gamma_0|)$. When $|\gamma_0| \ll t_0 \ll 1$ it yields considerable enhancement of superconductivity, $T_c\sim (1/\tau) e^{-2y_c}$, cf. Eq. \eqref{eq:N=1:Tc-RG}. We note that for $|\gamma_0| \ll t_0 \ll 1$ the attraction interaction $\gamma$ reaches value $t_0$ at the lengthscale $y_c-\ln 2$ and, then, very rapidly diverges. 

Next we tackle the case $\mathcal{N} = 0$ in the same manner. Strong spin-orbit coupling fully suppresses all triplet modes and equations for the remaining interactions $\gamma_s$, $\gamma_c$ along with resistivity $t$ read
\begin{subequations}
\begin{gather}
    \frac{dt}{dy} = - t^2/2 ,  \\
    \frac{d}{dy}\begin{pmatrix} \gamma_s \\ \gamma_c \end{pmatrix} = -\frac{t}{2} \begin{pmatrix} 1 & 2 \\ 1 & 0 \end{pmatrix} \begin{pmatrix} \gamma_s \\ \gamma_c \end{pmatrix} - \begin{pmatrix} 0 \\ 2\gamma_c^2 \end{pmatrix} . \label{eq:N=0:RG:a}
\end{gather}
\end{subequations}
Projection of the latter system onto the BCS line $-\gamma_s = \gamma_c = \gamma$ gives
\begin{equation}
    \frac{d\gamma}{dy} = (t/2)\gamma - (4/3)\gamma^2, \;\;\; \gamma_0 = (2\gamma_{c0}-\gamma_{s0})/3<0 .
\end{equation}
Similarly to the case discussed above, in the regime $|\gamma_0| \ll t_0 \ll 1$ 
$|\gamma|$ grows with increase of $y$ and reaches $t$. After this happen, $|\gamma|$ very quickly diverges at the length-scale $y_c \sim 1/\sqrt{|\gamma_0| t_0}$. The latter  corresponds to the following critical temperature $T_c\sim (1/\tau)e^{-2y_c}$, cf. Eq. \eqref{eq:N=0:Tc-RG} .

\section{Numerical solution}
\label{App-num}

In the current appendix we provide some details of the numerical solution for the critical temperature. For this purpose power iteration and dimensional fitting are employed. 

As mentioned in the main text, linearized self-consistency equation Eq. \eqref{eq:N=1:sc-linearized} can be thought of as an eigenvalue problem. However, the difficulty from a numerical point of view lies in the fact that the matrices defined in Eqs. \eqref{eq:N=1:sc-Tc} and \eqref{eq:N=0:sc-Tc} are parametrically large. Indeed, $n_{\max} \simeq 1/2\pi\tau T_c$ yields $n_{\max} \propto (t_0/|\gamma_0|)^{2/t_0} \gg 1 $ for $\mathcal{N} = 1$ and $n_{\max} \propto \exp(4/t_c - 4/t_0) \gg 1$ for $\mathcal{N} = 0$. Below we treat both cases separately.

When Ising spin-orbit coupling is considered, we fit the leading eigenvector with the expression 
\begin{equation}
    \lambda_M = c_1/t_0 + c_2,  \label{eq:AppB:N=1:fitting} 
\end{equation}
\noindent where $c_1$ and $c_2$ are the fitting parameters. This formula is justified by the following analytical estimate, 
\begin{gather}
    \lambda_M = \sum_{n' \geqslant 0}^{n_{\rm max}} \frac{\Delta_{n'}/\Delta_0}{(n'+1)^{t_0/2}(n'+1/2)} 
    \simeq \frac{2}{t_0} \int\limits_{u_\infty}^{u_0} \frac{du}{u_0} f(u) \notag \\ + c_2 = c_1/t_0 + c_2.
\end{gather}


\noindent This reveals the functional dependence of the leading eigenvalue $\lambda_M$ on the parameter $t_0 \ll 1$. In the above expression we have used the asymptotic expression for the right eigenvector $r_n$ in the form $r_n = f(u_n)$ and we have replaced $(n+n^\prime +1)$ with $\max\{(n+1/2)^{t_0/2}, (n^\prime+1/2)^{t_0/2}\}$, which is ensured by the smallness of the parameter $t_0$ and gives a correction of the order of $O(t_0)$. Fitting the numerical results with an analytical expression \eqref{eq:AppB:N=1:fitting}, we obtain $c_1\approx 1.38$ and $c_2 \approx 1.50$ which leads to the answer Eq. \eqref{eq:N=1:Tc-num}. 

It worth be pointed out that $\lambda_M \approx 1.38/t_0$ lies within the boundary of the Perron–Frobenius inequality,
\begin{gather}
    \lambda_M \leqslant \max\limits_{n} \sum_{n^\prime=0}^{n_{\max}} M_{nn^\prime} \simeq \frac{2}{t_0}. \label{eq:AppB:Tc-Eigenvalue} 
\end{gather}

Next we move on to the case $\mathcal{N} = 0$. Strong spin-orbit coupling corresponds to the matrix $M_{nn^\prime}(\ln n_{\max})$.
Again one can find the functional dependence of $\lambda_M$ on the matrix size $n_{\max}$ with the help of an approximation $\ln(n+n'+1)$ $\to$ $\ln(\max\{n+1/2, n'+1/2\})$. It reveals
\begin{gather}
     \lambda_M = \sum_{n'\geqslant 0}^{n_{\max}} \frac{\ln (n_{\max}) - \ln(n'+1)}{n'+1/2}  \frac{\Delta_{n'}}{\Delta_0} 
     \simeq \int\limits_{u_\infty}^{u_0} du u f(u) \notag \\ + c_2 \ln n_{\max}. \label{eq:AppB:N=0:estimation}
\end{gather}
With the help of Eqs. \eqref{eq:N=0:sc:Euler_Maclaurin} and \eqref{eq:N=0:Tc-EulerMaclaurin} and using $\pi^2/(|\gamma_0| t_0)=\ln^2 n_{\rm max}$, we find the following dimensional fitting,
\begin{equation}
    \lambda_M = c_1 \ln^2 n_{\max} + c_2 \ln n_{\max}.
\end{equation}
Using the above expression for the fit of numerical data we obtain $c_1 \approx 0.406$, $c_2 \approx 1.57$. Once again, one can check that $c_1 = 0.406$ satisfies the Perron-Frobenius inequality,
\begin{gather}
    \lambda_M \leqslant \max\limits_{n} \sum_{n^\prime=0}^{n_{\max}} M_{nn^\prime}(\ln n_{\max}) \simeq \frac{\ln^2 n_{\max}}{2}. 
\end{gather}

\section{Local density of states}
\label{App-LDoS}

In the present appendix we discuss details of calculation of the disorder-averaged pair correlation function of the local density of states. Within our approach the mesoscopic fluctuations of the local density of states can be expressed in terms of $Q$-matrices as follows, see \cite{BurmistrovPRL2013,BurmistrovPRB2015a},
\begin{gather}
    \notag K_2 (E, E^\prime, \bm{r})  = \langle \delta \rho(E, \bm{r})  \delta \rho(E^\prime, \bm{r}) \rangle 
     = \langle \rho(E, \bm{r}) \rho(E^\prime, \bm{r}) \rangle
     \notag \\ - \langle \rho(E, \bm{r}) \rangle \langle\rho(E^\prime, \bm{r}) \rangle  
    = \frac{\nu^2}{32} \re\big[ P_{2, irr}^{\alpha_1 \alpha_2}(i \varepsilon_1, i \varepsilon_3) \notag \\- P_{2, irr}^{\alpha_1 \alpha_2}(i \varepsilon_1, i \varepsilon_4) \big]. \label{eq:MFLDoS:pair-correlation}
\end{gather}
Here $\alpha_1 \neq \alpha_2$ are some fixed replica indices and analytical continuation, 
$i\varepsilon_{n_1} \to E + i0$,  $i\varepsilon_{n_3} \to E^\prime + i0$,  $i\varepsilon_{n_4} \to E^\prime - i0$, is assumed. $P_{2, irr}^{\alpha_1 \alpha_2}(i \varepsilon_n, i \varepsilon_m)$ is the irreducible part of bilinear in $Q$ operator,
\begin{equation}
 P_2^{\alpha_1 \alpha_2} = \langle \spp Q_{nn}^{\alpha_1 \alpha_1} \spp Q_{mm}^{\alpha_2 \alpha_2}  - 2 \spp Q_{nm}^{\alpha_1 \alpha_2} Q_{mn}^{\alpha_2 \alpha_1} \rangle.
\end{equation}
To find mesoscopic fluctuations of the local density of states, we approximate $P_{2, irr}^{\alpha_1 \alpha_2}(i\varepsilon_n,i\varepsilon_m)$ as 
\begin{gather}
P_{2, irr}^{\alpha_1 \alpha_2} \simeq -2 \spp\langle     
(R^{-1} W R)_{nm}^{\alpha_1 \alpha_2}
(R^{-1} W R)_{mn}^{\alpha_2 \alpha_1} \rangle .
\end{gather}
Using Eq. \eqref{eq:prop:Dp}, we find the following expression  
\begin{gather}
    \nonumber P_2(i\varepsilon, i\varepsilon^\prime) = -  \frac{32(1+\mathcal{N})}{g} \bigg[1 - \frac{\varepsilon}{\sqrt{\varepsilon^2 + \Delta^2}} \frac{\varepsilon^\prime}{\sqrt{\varepsilon^{\prime2} + \Delta^2}} \bigg] \\
    \times \int \frac{d^2 \bm{q}}{(2\pi)^2} \frac{D}{Dq^2 + \sqrt{\varepsilon^2+\Delta^2} + \sqrt{\varepsilon^{\prime2}+\Delta^2}},
\end{gather}
that works for all signs of $\varepsilon$ and $\varepsilon^\prime$. We point out that this expression is only valid under the assumption \eqref{eq:energy_range}, when energy dependence of the spectral gap function $\Delta_\varepsilon$ can be neglected. Then for $K_2(E, E^\prime, \bm{r})$, where $E, E^\prime \gg \Delta$, we find the following lengthy expression, 
\begin{gather}
   K_2 = \frac{\nu^2(1+\mathcal{N})}{g} \sum_{s=\pm} s  
  \left(1 + \frac{s E E^\prime}{\sqrt{E^2-\Delta^2}\sqrt{E^{\prime 2}-\Delta^2}} \right) \notag\\
     \times\re \int_q \frac{D}{Dq^2 + i\sqrt{E^2-\Delta^2} -i s \sqrt{E^{\prime 2} - \Delta^2 }} .  \label{eq:AppD:K2}
\end{gather}

\noindent Clearly, when $E = E^\prime$, we obtain Eq. \eqref{eq:varLDoS:1}. 

We proceed to mesoscopic fluctuations of the local density of states at different energies $E \neq E^\prime$. First of all, let us note that when energies $E$ and $E^\prime$ are close a pair correlation function $K_2(E, E^\prime, \bm{r})$ differs only slightly from the one on coinciding energies, $K_2(E, E, \bm{r})$. Therefore the case of energies separated far away from one another is of the interest. Assume that $E^\prime = E+\omega$, where $\omega$ is large in comparison to $E$. Then, expanding the \eqref{eq:AppD:K2} in a small argument $E/\omega \ll 1$ and performing the integration with respect to momentum $\bm{q}$ we obtain Eq. \eqref{eq:MFLDoS:K2:E(E+w)}.


\end{document}